\documentclass[11pt]{article}
\baselineskip
\usepackage{amssymb}
\usepackage{graphicx}
\usepackage{latexsym,cite}
\usepackage{epsfig}
\usepackage{amsmath, bbm, color}

\newcommand{\SU}{\mathrm{SU}}
\newcommand{\U}{\mathrm{U}}
\newcommand{\Lambdamsbar}{\Lambda_{\overline{\rm MS}}}
\newcommand{\eq}[1]{\begin{equation}\label{#1}}
\newcommand{\en}{\end{equation}}
\newcommand{\eqar}[1]{\begin{eqnarray}\label{#1}}
\newcommand{\enar}{\end{eqnarray}}

\begin{document}

\begin{titlepage}
\vskip0.5cm
\begin{center}
{\Large\bf Thermodynamics of the QCD plasma and the large-$N$ limit}
\end{center}
\vskip1.3cm
\centerline{Marco~Panero$^{a, b}$}
\vskip1.5cm
\centerline{\sl  $^a$ Institute for Theoretical Physics,  ETH Z\"urich, 8093 Z\"urich, Switzerland}
\vskip0.5cm
\centerline{\sl  $^b$ Institute for Theoretical Physics,  University of Regensburg, 93040 Regensburg, Germany}
\vskip0.5cm
\begin{center}
{\sl  e-mail:} \hskip 6mm \texttt{panero@phys.ethz.ch}
\end{center}
\vskip1.0cm
\begin{abstract}
The equilibrium thermodynamic properties of the $\SU(N)$ plasma at finite temperature are studied non-perturbatively in the large-$N$ limit, via lattice simulations. We present high-precision numerical results for the pressure, trace of the energy-momentum tensor, energy density and entropy density of $\SU(N)$ Yang-Mills theories with $N=3$, $4$, $5$, $6$ and $8$ colors, in a temperature range from $0.8T_c$ to $3.4T_c$ (where $T_c$ denotes the critical deconfinement temperature). The results, normalized according to the number of gluons, show a very mild dependence on $N$, supporting the idea that the dynamics of the strongly-interacting QCD plasma could admit a description based on large-$N$ models. We compare our numerical data with general expectations about the thermal behavior of the deconfined gluon plasma and with various theoretical descriptions, including, in particular, the improved holographic QCD model recently proposed by Kiritsis and collaborators. We also comment on the relevance of an AdS/CFT description for the QCD plasma in a phenomenologically interesting temperature range where the system, while still strongly-coupled, approaches a `quasi-conformal' regime characterized by approximate scale invariance. Finally, we perform an extrapolation of our results to the $N \to \infty$ limit.
\end{abstract}
\vspace*{0.2cm}

\end{titlepage}

\section{Introduction}
\label{sec:intro}

A direct implication of asymptotic freedom in non-Abelian gauge theories~\cite{Gross:1973id, Politzer:1973fx} is that hadronic matter should undergo a change of state to a deconfined phase at sufficiently high temperatures or densities~\cite{Cabibbo:1975ig, Collins:1974ky}. This theoretical prediction has been subject to extensive experimental investigation through heavy-ion collisions for more than two decades. The evidence obtained from the SPS and RHIC experiments led to the conclusion that `a new state of matter' was created~\cite{Heinz:2000bk, Adcox:2004mh, Arsene:2004fa, Back:2004je, Adams:2005dq}, which reaches rapid thermalization, and is characterized by very low viscosity values, making it a nearly ideal fluid~\cite{Kolb:2003dz}; a brief summary of experimental results can be found in ref.~\cite{BraunMunzinger:2007zz}.

On the theoretical side, however, the understanding of such strongly-interacting quark-gluon plasma (sQGP) is still an open issue~\cite{Shuryak:2008eq}. Relativistic fluidodynamics---see ref.~\cite{Romatschke:2009im} for an introduction to the subject---provides a successful phenomenological description of the experimental data, but it is not derived from the first principles of QCD. On the other hand, the weak-coupling expansion in finite-temperature QCD is non-trivial: it contains terms proportional to odd powers of the gauge coupling~\cite{Kapusta:1979fh}, as well as contributions from diagrams involving arbitrarily large numbers of loops~\cite{Linde:1980ts, Gross:1980br}, and is slowly convergent; finally, it shows quite large discrepancies with the results determined non-perturbatively from lattice simulations at temperatures close to the deconfinement~\cite{Arnold:1994ps, Arnold:1994eb, Zhai:1995ac, Kajantie:2002wa, Laine:2005ai, Hietanen:2008tv}. One possibility is that this mismatch is due to intrinsically non-perturbative contributions proportional to $T^2$~\cite{Pisarski:2006hz, Pisarski:2006yk}, which could perhaps be modelled in terms of a modified bag model~\cite{Chodos:1974je} with finite thickness; related ideas have also been discussed in refs.~\cite{Megias:2003ui, Megias:2005ve, Andreev:2007zv, Megias:2009mp, Megias:2009ar}. Alternative descriptions of the sQGP phenomenology are based on various types of effective degrees of freedom~\cite{Blaizot:2001nr, Kraemmer:2003gd, Shuryak:2008eq, Antonov:2008kb, Brau:2009mp, Buisseret:2009eb}. 

For a first-principle description of the QCD plasma, however, the numerical approach on the lattice is essentially the main tool. In particular, the lattice determination of equilibrium thermodynamic observables in the $\SU(3)$ gauge theory is considered as a solved problem~\cite{Boyd:1996bx}; on the other hand, several studies of finite-temperature QCD with dynamical fermions have begun appearing during the last decade~\cite{Karsch:2000ps, AliKhan:2001ek, Aoki:2005vt, Bernard:2006nj, Cheng:2007jq, Bazavov:2009zn, Aoki:2009sc, Cheng:2009zi}, and today large-scale unquenched simulations can be done at or close to the physical point; recent results are reviewed in refs.~\cite{DeTar:2008qi, DeTar:2009ef, Laine:2009ik}. 

Yet, studies of the thermodynamics of the pure-gauge sector can still be relevant from a fundamental perspective, as they capture the essential qualitative features of the deconfinement phenomenon, and offer several advantages. 

Firstly, the deconfining transition at finite temperature in pure Yang-Mills (YM) models is characterized by a well-defined theoretical setup, with a clean symmetry pattern: in the Euclidean formulation, the action is invariant under center symmetry, and the trace of the Wilson line (or Polyakov loop) along the compactified Euclidean time direction serves as the order parameter distinguishing the confined (symmetric) from the deconfined phase (with spontaneously broken symmetry). By contrast, in full QCD with physical quark masses the center symmetry is explicitly broken by the quarks, and no unambiguous definition of a critical temperature exists, with the change of state from the confined to the deconfined phase being a cross-over. 

Secondly, calculations in lattice QCD including fermionic degrees of freedom are computationally much more demanding than in YM models, and all presently known methods to implement the fermionic sector on the lattice have advantages and drawbacks~\cite{Jansen:2008vs}. On the other hand, numerical simulations in the pure gauge sector can attain much higher precision, and allow to detect fine details in the behavior of various observables.

Finally, the study of $\SU(N)$ gauge models with $N \ge 3$ colors is relevant for the large-$N$ limit. Historically, the idea of studying QCD in the large-$N$ limit was first put forward in 1974 by 't~Hooft~\cite{'tHooft:1973jz}, who proposed to consider $1/N$ as an expansion parameter: the limit for $N$ going to infinity at fixed bare 't~Hooft coupling $\lambda_0=g_0^2 N$ ($g_0$ being the bare YM coupling) yields a simpler theory, characterized by dominance of a subset of Feynman diagrams (which, in a double-line notation, can be represented as planar ones), and amenable to a topological expansion in terms of surfaces of increasing genus. Indeed, the simplifications taking place in the large-$N$ model can be useful to obtain insight into certain non-trivial non-perturbative features of QCD~\cite{Witten:1979kh, Manohar:1998xv}. As finite-$N$ corrections are of order $\mathcal{O}(N^{-1})$ for QCD (or $\mathcal{O}(N^{-2})$ for pure YM), one may expect that predictions derived from large-$N$ QCD have phenomenological relevance; several issues related to this expectation have been investigated on the lattice by different groups---see refs.~\cite{Teper:2008yi, Narayanan:2005en, Vicari:2008jw} and references therein. 

However, from a more modern perspective, the large-$N$ limit has much farther-reaching implications (see, \emph{e.g.}, refs.~\cite{Unsal:2008ch, Zhitnitsky:2008ha} and references therein), and also entails deep connections with string theory. Indeed the large-$N$ limit is one of the crucial ingredients underlying the AdS/CFT correspondence conjectured by Maldacena~\cite{Maldacena:1997re}, according to which the large-$N$ limit of the maximally supersymmetric $\mathcal{N}=4$ supersymmetric YM (SYM) theory in four dimensions is dual to type IIB string theory in a $AdS_5 \times S^5$ space. In principle, this highly non-trivial correspondence allows one to study the strongly-coupled regime of field theory by mapping it to the weak-coupling limit of a gravity model. The AdS/CFT correspondence is an example of a duality relating the large-$N$ limit of gauge theories with $\U(N)$ structure group and string theory. The mathematical meaning of such dualities can be summarized in the following terms: gauge theories deal with invariants originating from moduli spaces of connections on principal bundles, whereas string theory studies the invariants arising from spaces of maps from certain classes of domains onto different targets; large-$N$ dualities relate the generating functions of such invariants, \emph{i.e.} the partition functions of these theories~\cite{Auckly:2007zw}.

Although the Maldacena conjecture was originally formulated for the $\mathcal{N}=4$ SYM theory, the techniques underlying the AdS/CFT correspondence are often applied to theories which are expected to have properties more similar to QCD: in the so-called `top-down' approach, this is achieved by explicitly breaking some symmetries of the $\mathcal{N}=4$ theory. This allows to build models with a non-trivial hadron sector, which can be studied non-perturbatively using the dual gravity formulation~\cite{Erdmenger:2007cm}; for the meson sector, the predictions obtained from these calculations have been compared with (quenched) spectroscopy calculations in $\SU(N)$ lattice gauge models~\cite{DelDebbio:2007wk, Bali:2008an}, finding satisfactory agreement. Another field where analytical techniques inspired by the AdS/CFT correspondence are applied is in the description of hydrodynamic and thermodynamic properties relevant for a strongly interacting system, like the QCD plasma---see refs.~\cite{Policastro:2001yc, Herzog:2006gh, Liu:2006ug, Gubser:2006bz, Son:2007vk, Mateos:2007ay, Gubser:2009md, Brasoveanu:2009ky, Rangamani:2009xk} and references therein. In principle, the appropriate parameters for the gravity models dual to the sQGP can be pinned down using the combined constraints coming from experimental data and lattice QCD simulations~\cite{Noronha:2009vz}.

On the other hand, similar analytical techniques have also been used in a `bottom-up' approach, building five-dimensional holographic models that should reproduce the non-perturbative features of QCD~\cite{Polchinski:2001tt, Erlich:2005qh, DaRold:2005zs, Karch:2006pv}; this approach has proven quite successful for the hadron properties, but is not completely satisfactory for the QCD plasma thermodynamics~\cite{Herzog:2006ra}. More recently, a related approach has been used to build an ``improved holographic QCD model''~\cite{Gursoy:2007cb, Gursoy:2007er, Kiritsis:2009hu, Gursoy:2008bu} through a dual description in terms of a dilaton-gravity system in five dimensions; this model reproduces the main non-perturbative features of the $\SU(3)$ YM spectrum as determined in lattice calculations, as well as the equilibrium thermodynamic properties of the sQGP~\cite{Gursoy:2008bu, Gursoy:2008za, Gursoy:2009jd}; furthermore, the model can also be used to derive predictions for observables which are more difficult to study on the lattice, such as the plasma bulk viscosity, drag force and jet quenching parameter~\cite{Gursoy:2009kk}. Related works include refs.~\cite{Nojiri:1999gf, Andreev:2006vy, Csaki:2006ji, Gubser:2008ny, Gubser:2008yx, Evans:2008tu, Batell:2008zm, dePaula:2008fp, Noronha:2009ud, Alanen:2009xs}.

Clearly, all models aiming at a description of $N=3$ QCD (or YM) in terms of predictions based on the large-$N$ limit implicitly rely on the assumption that the features of the $N=3$ theory are `close enough' to those of its $N=\infty$ counterpart. While \emph{a priori} this assumption is not guaranteed to be true, there is strong numerical evidence that this is indeed the case~\cite{Teper:2008yi}. In this context, the aim of the work reported here was to study non-perturbatively the equilibrium thermodynamics properties at finite temperature in $\SU(N)$ YM theories with $N=3$, $4$, $5$, $6$ and $8$ colors, via lattice simulations. Related studies include works by Bringoltz and Teper~\cite{Bringoltz:2005rr, Bringoltz:2005kx} and by Datta and Gupta~\cite{Datta:2009tj, Datta:2009jn, Datta:2009ef}.

The structure of this paper is the following: in sec.~\ref{sec:theory}, we briefly review some theoretical expectations for the sQGP thermodynamics, which can be compared with lattice results at large $N$. The setup of our lattice calculation is formulated in sec.~\ref{sec:lattice}, while in sec.~\ref{sec:results} we present the numerical results for the pressure, trace of the stress-energy tensor, energy and entropy densities, and compare them with the theoretical expectations; we discuss the main uncertainties involved in our calculation, and the extrapolations to the continuum and to the large-$N$ limit. Finally, we conclude in sec.~\ref{sec:conclusions} with a summary of our findings, a discussion about their implications, and an outlook on future work. Preliminary results of this study have been presented in ref.~\cite{Panero:2008mg}.

\section{Pictures of the Unicorn: Theoretical descriptions of the sQGP thermodynamics}
\label{sec:theory}

In the literature, there exist many theoretical approaches to describe the sQGP thermodynamics; the elusiveness of the QCD plasma, which has inspired the comparison with the mythological creature (see ref.~\cite{Pisarski:2006ie} and ref.~[3] therein), is not only a matter of experimental issues, but is also related to the theoretical difficulties in working out a quantitatively successful description for a strongly-interacting system. 

While a review of the theoretical approaches would be a task beyond the scope of this paper (we refer the reader to reviews available in the literature, like, \emph{e.g.}, refs.~\cite{Blaizot:2001nr, Shuryak:2008eq}), in this section we just recall certain theoretical expectations, for which a comparison with the numerical results of $N \ge 3$ simulations may be of interest. 

Roughly speaking, the logical organization of this section is from the perturbative to the non-perturbative, from QCD to strings, from lower to higher dimensions. More precisely, we start by briefly discussing the key issues in the weak-coupling expansion for thermal QCD, and the possibility to resolve its mismatch with lattice results at temperatures of the order of the deconfinement temperature $T_c$ by including terms of non-perturbative origin, which could perhaps be described in terms of a ``fuzzy bag model''. Then we move to the description of the sQGP by gravity models, reviewing the basic features of the improved holographic QCD model, and recalling some AdS/CFT predictions which can be compared with our numerical results.

\subsection{Weak-coupling expansions in hot QCD}
\label{subsec:weak_coupling}

Due to asymptotic freedom, the weak-coupling expansion is expected to be a natural analytical approach to the thermodynamics of hot QCD~\cite{Kraemmer:2003gd}. In particular, at very high temperatures QCD develops a hierarchy of momentum scales, with a clear separation between `hard', `soft' and `ultrasoft' modes, whose parametric dependence on the temperature $T$ and on the coupling $g$ is $\mathcal{O}(T)$, $\mathcal{O}(gT)$ and $\mathcal{O}(g^2T)$, respectively. By integrating out the hard modes, one can derive a dimensionally-reduced effective theory~\cite{Ginsparg:1980ef, Appelquist:1981vg, Braaten:1995jr, Braaten:1995cm}, whereby the QCD pressure can be studied in a weak-coupling expansion~\cite{Kajantie:2000iz, Kajantie:2002wa, DiRenzo:2008en} and some of the intrinsically non-perturbative terms~\cite{Linde:1980ts, Gross:1980br} can be estimated numerically~\cite{Hietanen:2004ew, Hietanen:2006rc, DiRenzo:2006nh, Hietanen:2008tv}. However, the convergence of the perturbative series appears to be rather slow, and the most recent results (for the pure gauge sector)~\cite{Hietanen:2008tv} exhibit large deviations from the corresponding lattice calculations at temperatures $\mathcal{O}(T_c)$~\cite{Boyd:1996bx}; perhaps the situation could be improved using a center-symmetric  dimensionally-reduced theory~ \cite{Vuorinen:2006nz, deForcrand:2008aw, Kurkela:2007dh}.

\subsection{Non-ideal contributions to the QCD pressure and a ``fuzzy bag'' model}
\label{subsec:fuzzy_bag_model}

The mismatch between the weak-coupling expansion for the pressure in thermal QCD and lattice results is particularly severe for $ T_c \le T \le 3 T_c$, see, \emph{e.g.}, fig.~6 in ref.~\cite{Hietanen:2008tv}. Pisarski observed that in this temperature range the lattice results for the trace of the stress-energy tensor (often denoted as $\Delta$) in the $\SU(3)$ gauge theory appear to be well-described by a $T^2$ behavior~\cite{Pisarski:2006hz, Pisarski:2006yk}, which is not captured by the weak-coupling expansion. For this reason, he argued that the expression of the pressure $p$, as a function of the temperature in the range between the maximum of $\Delta/T^4$ (which, for $\SU(3)$, is located around $1.1 T_c$) and the onset of the perturbative behavior, should be of the form~\cite{Pisarski:2006yk}:
\eq{Pisarski_pressure}
p(T) = f_{\mbox{\tiny{pert}}} T^4 - B_{\mbox{\tiny{fuzzy}}} T^2 - B_{\mbox{\tiny{MIT}}} + \dots
\en
where $f_{\mbox{\tiny{pert}}}$ includes the perturbative contributions, $B_{\mbox{\tiny{fuzzy}}}$ is a ``fuzzy bag'' term, and $B_{\mbox{\tiny{MIT}}}$ is the term associated with the usual MIT bag model~\cite{Chodos:1974je}. The idea behind the $B_{\mbox{\tiny{fuzzy}}}$ term is that of a QCD vacuum where deconfined ``bubbles'' have a surface of finite thickness; if the degrees of freedom associated with fluctuations of this thick bag surface are collective ones and involve the whole surface width, then they are expected to be difficult to excite, thus they should not lead to (unphysical) multiplicities in the zero-temperature hadron spectrum. On the other hand, assuming that the surface of the fuzzy bag extends over a width between $0.3$~fm and the typical hadron size of $1$~fm would make it plausible that it should become relevant for temperatures in a range between $T_c$ and $3T_c$. 

From this point of view, one can consider the fuzzy bag model as a phenomenological description for the QCD vacuum and for the plasma thermodynamics; such description involves two free parameters ($B_{\mbox{\tiny{fuzzy}}}$ and $B_{\mbox{\tiny{MIT}}}$). One possible way to relate this description to more fundamental degrees of freedom is in terms of an effective theory of Wilson lines, whose dynamics could be described by some random matrix model~\cite{Guhr:1997ve, Caselle:2003qa}; in this respect, confinement would arise from the eigenvalue repulsion in the large-$N$ limit---an issue which has been studied
in a finite volume in ref.~\cite{Aharony:2005bq}.

\subsection{Thermodynamics from the improved holographic QCD model}
\label{subsec:improved_holographic_QCD}

A radically different description for the YM spectrum and for the deconfined gluon plasma can be obtained from the improved holographic QCD (IHQCD) model proposed by Kiritsis and collaborators~\cite{Gursoy:2007cb, Gursoy:2007er, Kiritsis:2009hu, Gursoy:2008bu, Gursoy:2008za, Gursoy:2009jd, Gursoy:2009kk}, which is an AdS/QCD model based on an Einstein-dilaton gravity theory in five dimensions, where the fifth (radial) direction is dual to the energy scale of the $\SU(N)$ gauge theory. On the gravity side, the bulk fields are the metric, the dilaton and the axion, which are respectively dual to the energy-momentum tensor, to the trace of $F^2$, and to the trace of $F \tilde F$ on the gauge side. The action is defined as:
\eq{IHQCD_action}
S_{IHQCD} = -M_P^3 N^2\int d^5x\sqrt{g}
\left[R-{4\over 3}(\partial\Phi)^2+V(\lambda) \right]+2M_P^3 N^2\int_{\partial M}d^4x \sqrt{h}~K,
\en
where $\Phi$ is the dilaton field, while $\lambda=\exp(\Phi)$ is identified with the running 't~Hooft coupling of the dual $\SU(N)$ YM theory, $M_P$ denotes the five-dimensional Planck scale, and $K$ is the extrinsic curvature of the boundary \emph{\`a la} Gibbons-Hawking. Note that the effective five-dimensional Newton constant is $G_5 = 1/\left(16 \pi M_P^3 N^2\right)$, which becomes small in the large-$N$ limit. The dilaton potential $V(\lambda)$ is defined according to the requirements of asymptotic freedom with a logarithmically running coupling and linear confinement in the ultraviolet (UV) and infrared (IR) limits of the dual gauge model; one possible \emph{Ansatz} is:
\eq{IHQCD_potential}
V(\lambda)= \frac{12}{\ell^2} \left[ 1 + V_0 \lambda + V_1 \lambda^{4/3} \sqrt{ \log \left(1 + V_2 \lambda^{4/3} + V_3 \lambda^2\right) } \right] ,
\en
where $\ell$ is the AdS scale (which only defines the overall normalization of the potential). $V_0$, $V_1$, $V_2$ and $V_3$ are free parameters: two of them can be fixed by imposing that the dual model reproduces the first two (scheme-independent) perturbative coefficients of the $\SU(N)$ $\beta$-function, and one is left with two independent parameters. The latter can be adjusted by requiring that the values for two independent physical quantities in the dual model match the corresponding YM results obtained from lattice simulations.

In this model, the first-order transition from a confined to a deconfined phase (respectively dual to a thermal-graviton- and to a black-hole-dominated regime of the five-dimensional gravity theory) is associated with the appearance of a non-trivial gluon condensate. 

Strictly speaking, the IHQCD model is expected to hold in the large-$N$ only, while at finite $N$ one expects corrections; in particular, the calculations in the gravity model neglect string interactions, which are expected to become important above a cut-off scale $M_P N^{2/3}$; for the values of the parameters used in ref.~\cite{Gursoy:2009jd}, this would approximately correspond to $2.5$~GeV in $\SU(3)$---see also ref.~\cite{Gursoy:2009zza}.

\subsection{A prediction from $\mathcal{N}=4$ SYM thermodynamics: The entropy deficit}
\label{subsec:N=4_SYM}

While the IHQCD model discussed in subsec.~\ref{subsec:improved_holographic_QCD} provides a quantitatively accurate \emph{description}---rather than a \emph{prediction}---for the non-perturbative features of the strongly-interacting YM spectrum and of the deconfined plasma, and is based on a non-supersymmetric, non-critical string formulation~\cite{Polyakov:1998ju, Klebanov:2004ya}, a more fundamental/less phenomenological approach consists in deriving general properties of the sQGP directly from the AdS/CFT setup---see refs.~\cite{Mateos:2007ay, Son:2007vk, Gubser:2009md, Brasoveanu:2009ky, Rangamani:2009xk} and references therein. 

It is worth stressing that, in principle, there are no obvious reasons to expect the large-$N$ limit of $\mathcal{N}=4$ SYM to be a theory ``close'' to real-world QCD: as opposed to the latter, the former is maximally supersymmetric, conformally invariant, lacks confinement and spontaneous breaking of chiral symmetry. The field content of the two theories is also very different, given that SYM features fermions in the adjoint, rather than in the fundamental, representation, and that the number of colors $N$ is infinite, rather than $3$ as in QCD.

Yet, this approach has led to the analytical derivation of important results, with potential relevance also for phenomenology. A celebrated example is the value for the shear viscosity to entropy density ratio $\eta/s=1/(4\pi)$~\cite{Policastro:2001yc}, which is expected to hold for all gauge theories in the limit of infinite 't~Hooft coupling~\cite{Buchel:2003tz}, and which is conjectured to be a universal lower bound for all substances existing in Nature~\cite{Kovtun:2004de}.

In the context of the equilibrium thermodynamics quantities that we studied in the present work, one important prediction from large-$N$ $\mathcal{N}=4$ SYM concerns the entropy density $s$, which can be evaluated in the limit of infinite coupling $\lambda \to \infty$ by identifying it with the Bekenstein-Hawking entropy density of the dual gravity theory: the result is found to be equal to $3/4$ of the entropy density calculated in the limit of zero coupling (denoted as $s_0$)~\cite{Gubser:1996de}. This result was later refined in ref.~\cite{Gubser:1998nz}, with the evaluation of the leading correction at large but finite $\lambda$ values:
\eq{sugra_entropy}
\frac{s}{s_0} = \frac{3}{4} + \frac{45}{32} \zeta(3) ( 2 \lambda )^{-3/2} + \dots \nonumber
\en
where $\zeta$ is the Riemann function, and the dots denote subleading corrections; as Ap\'ery's constant $\zeta(3) \simeq 1.2020569$ is a positive number, the asymptotic value of $s/s_0$ is approached from above when $\lambda$ tends to infinity.

Note that---as opposed to QCD or to the IHQCD model discussed in subsec.~\ref{subsec:improved_holographic_QCD}---in the supersymmetric model $\lambda$ is merely a parameter of the theory, which does not run with the energy.

\section{Lattice formulation}
\label{sec:lattice}

In this work, we performed a non-perturbative study of $\SU(N)$ YM theories with $N=3$, $4$, $5$, $6$ and $8$ colors, regularizing them on a four-dimensional Euclidean hypercubic, isotropic lattice with periodic boundary conditions in all directions, and estimating the expectation values of physical observables via Monte~Carlo simulations. In the following, $a$ denotes the lattice spacing, $N_s$ ($N_t$) is the number of lattice points along each of the three space-like (along the time-like) directions, while $V=(aN_s)^3$ and $T=1/(aN_t)$ denote the spatial volume and temperature, respectively. The degrees of freedom of the theory are represented as $U_\mu(x) \in \SU(N)$ matrices of size $N \times N$, defined on the oriented elementary bonds $(x,\mu)$ of the lattice; their dynamics is governed by the standard (unimproved) Wilson gauge action $S_W$~\cite{Wilson:1974sk}:
\eq{wilsonaction}
S_W = \sum_p \left( 1 - \frac{1}{N} \mbox{ReTr} U_p \right) ,
\en
where the summation ranges over all the elementary lattice plaquettes $p$, with $U_p$ denoting the path-ordered product of the link variables around each plaquette. The partition function $Z=Z(T,V)$ is given by:
\eq{partitionfunction}
Z(T,V) = \int D U \exp ( - \beta S_W ) ,
\en
where $ D U $ denotes integration over the Haar measure for all $U_\mu(x)$, and $\beta$ is related to the bare lattice gauge coupling $g_0$ through: $\beta=2N/g_0^2$. The expectation values of the various observables are estimated via a dynamic Monte Carlo simulation sampling the space of configurations of the system. The configuration update algorithm to generate the Markov chain is based on a combination of heat-bath steps for the $\SU(2)$ subgroups of $\SU(N)$~\cite{Creutz:1980zw, Kennedy:1985nu, Cabibbo:1982zn} and full $\SU(N)$ overrelaxations~\cite{Kiskis:2003rd, deForcrand:2005xr}---see also ref.~\cite{Durr:2004xu}---, except for a subset of the $T=0$ simulations, which were performed using the Chroma library~\cite{Edwards:2004sx} on the QCDOC machine~\cite{Boyle:2003ue} in Regensburg. 

The physical scale for the $\SU(3)$ gauge group was set using the values of Sommer's parameter $r_0$~\cite{Sommer:1993ce} taken from ref.~\cite{Necco:2001xg}. For the other $\SU(N>3)$ groups, the physical scale was defined from interpolation of high-precision measurements of the string tension $\sigma$~\cite{Lucini:2004my, Lucini:2000qp}; for the $\beta$ values beyond the range where high-precision string tension measurements are available, this was combined with a three-loop perturbative expansion~\cite{Alles:1996cy} in terms of the Parisi mean-field improved bare lattice coupling~\cite{Parisi:1980pe}, taking lattice corrections into account, following ref.~\cite{Allton:2008ty}.

The parameters of the main set of lattice simulations performed in this work (denoted as `ensemble A' in the following) are summarized in tab.~\ref{tab:ensemble_A_simulation_info}

\begin{table}[h]
\centering
\begin{tabular}{|c|cccccc|ccc|}  
\hline
$N$ & $N_s$ & $N_t$ & $\beta_{\mbox{\tiny{min}}}$ & $\beta_{\mbox{\tiny{max}}}$ & $N_\beta$ & $\Delta \beta$ & $\beta$-range & $T=0$ statistics & finite-$T$ statistics \\
\hline \hline
$3$ & $20$ & $5$ & $5.7$ & $6.6$ & $181$ & $0.005$ &  
$[5.7,5.795]$ & $3000$ & $12000$ \\
 & \multicolumn{6}{|c|}{ } & $[5.8,5.845]$ & $5800$ & $33600$ \\
 & \multicolumn{6}{|c|}{ } & $[5.85,5.895]$ & $5800$ & $21600$ \\
 & \multicolumn{6}{|c|}{ } & $5.9$ & $6400$ & $21600$ \\
 & \multicolumn{6}{|c|}{ } & $[5.905,6.2]$ & $5800$ & $21600$ \\
 & \multicolumn{6}{|c|}{ } & $[6.205,6.6]$ & $3000$ & $12000$ \\
\hline
$4$ & $16$ & $5$ & $10.5$ & $11.8$ & $131$ & $0.01$ &  
$[10.5,11.5]$ & $16000$ & $24000$ \\
 & \multicolumn{6}{|c|}{ } & $[11.51,11.67]$ & $4200$ & $12300$ \\
 & \multicolumn{6}{|c|}{ } & $11.68$ & $7400$ & $12300$ \\
 & \multicolumn{6}{|c|}{ } & $[11.69,11.8]$ & $4200$ & $12300$ \\
\hline
$5$ & $16$ & $5$ & $16.75$ & $18.45$ & $35$ & $0.05$ &  
$[16.75,18.45]$ & $2500$ & $7500$ \\
\hline
$6$ & $16$ & $5$ & $24.4$ & $26.9$ & $26$ & $0.1$ &  
$[24.4,25.5]$ & $10000$ & $20000$ \\
 & \multicolumn{6}{|c|}{ } & $[25.6,26.9]$ & $5200$ & $8400$ \\
\hline
$8$ & $16$ & $5$ & $43.9$ & $48.1$ & $43$ & $0.1$ &  
$[43.9,48.1]$ & $3200$ & $9200$ \\
\hline
\end{tabular}
\caption{Parameters of the main set of lattice simulations used in this work (ensemble A). $N$ denotes the number of colors, $N_s$ and $N_t$ are the space-like and, respectively, time-like sizes of the system in lattice units, $N_\beta$ is the number of $\beta$-values that were simulated, separated by $\Delta \beta$ from each other, in the $\beta_{\mbox{\tiny{min}}} \le \beta \le \beta_{\mbox{\tiny{max}}}$ interval. For each subset of $\beta$-values in this range, the $T=0$ and finite-$T$ statistics at each $\beta$ are shown in the last two columns.}\label{tab:ensemble_A_simulation_info}
\end{table}

We focus on the following equilibrium thermodynamic quantities: free energy density ($f$), pressure ($p$), energy density ($\epsilon$), trace of the energy-momentum tensor ($\Delta$) and entropy density ($s$):
\eqar{definitions}
f & = & - \frac{T}{V} \log Z \\
p & = & T \left. \frac{\partial \log Z}{\partial V} \right|_T \\
\epsilon & = & \frac{T^2}{V} \left. \frac{\partial \log Z}{\partial T} \right|_V \\
\Delta &=& \epsilon - 3p \\
s &=&  \frac{1}{V} \log Z +\frac{\epsilon}{T}.
\enar
In the thermodynamic limit one has: $p=-f$, while in a finite hypertorus in general this relation gets infrared corrections~\cite{Engels:1981ab, Elze:1988zs, Kapusta:1989cr, Gliozzi:2007jh, Campos:2008df, Meyer:2009kn}. In particular, the infrared corrections to the equation of state of a gas of free gluons in a finite volume were derived analytically in ref.~\cite{Gliozzi:2007jh}; such corrections could be particularly relevant for lattice simulations at very high temperatures~\cite{Endrodi:2007tq}, especially for the case of a fixed value of the aspect ratio of a time-like cross-section of the system $LT$. Although the investigation of such a limit (\emph{i.e.} at a very fine lattice spacing $a$, with a shrinking physical space-like volume) is interesting in its own~\cite{Bruckmann:2008xr, Bruckmann:2008rj}, in the present work we restricted our attention to relatively low temperatures, in a range up to about $3T_c$, where---due to screening~\cite{Gross:1980br}---such effects are not visible within the level of precision of our simulations. This was investigated in our preliminary study reported in ref.~\cite{Panero:2008mg}, which showed that, for the setup of the calculations presented in this work, finite-size effects are negligible.

In the regime where finite-volume corrections can be neglected, the pressure---relative to its $T=0$ vacuum value---can be conveniently estimated by means of the ``integral method''~\cite{Engels:1990vr}:
\eq{integral_method}
p \simeq \frac{T}{V} \log Z = \frac{1}{a^4 N_s^3 N_t} \int_{\beta_0}^{\beta} d \beta^\prime \frac{\partial \log Z}{\partial \beta^\prime},
\en
where $\beta_0$ is chosen to be a point sufficiently deep in the confined phase (corresponding to a temperature at which the pressure differs from its $T=0$ value by a negligible amount).

The dimensionless ratio $p/T^4$ can be readily evaluated from:
\eq{lattice_pressure}
\frac{p}{T^4} = 6 \int_{\beta_0}^{\beta} d \beta^\prime \left( \langle U_p \rangle_T - \langle U_p \rangle_0 \right),
\en
where the notation $\langle \dots \rangle_T$ indicates the expectation value in the system at temperature $T$ (which is estimated from simulations on a lattice of sizes $N_s^3 \times N_t$), while $\langle \dots \rangle_0$ denotes the expectation value at zero temperature (that can be estimated from measurements on a lattice of size $N_s^4$).

The integration in eq.~(\ref{lattice_pressure}) was performed numerically, dividing the integration range in $\beta$ in $N_\beta-1$ equally spaced intervals (see tab.~\ref{tab:ensemble_A_simulation_info}) and according to the methods discussed in the Appendix of ref.~\cite{Caselle:2007yc}.

$\Delta/T^4$ can be expressed as:
\eq{lattice_rescaled_trace}
\frac{\Delta}{T^4} = 6 N_t^4 
\left( \langle U_p \rangle_0 - \langle U_p \rangle_T \right)
\cdot a \frac{\partial \beta}{\partial a}.
\en

The energy and entropy densities can then be obtained indirectly, as linear combinations of $p/T^4$ and $\Delta/T^4$, namely:
\eq{lattice_energy_density}
\frac{\epsilon}{T^4}=\frac{\Delta+3p}{T^4}
\en
and:
\eq{lattice_entropy_density}
\frac{s}{T^3}=\frac{\Delta+4p}{T^4}.
\en

Finally, it should be noted that lattice cut-off effects induce a systematic deviation from the continuum values of these quantities. In the infinite-temperature limit, such effect is quantified by a dimensionless ratio denoted as $R_I(N_t)$, which can be evaluated by numerical integration~\cite{Engels:1999tk}. In the present work, we always include this factor in the definition of the lattice Stefan-Boltzmann (SB) limit for the various thermodynamic observables.

\section{Results}
\label{sec:results}

In this section, we present the numerical results from our main set of simulations on lattices with $N_t=5$ (ensemble A) and compare them with the theoretical predictions in subsection~\ref{subsec:Nt5_results}. The systematic uncertainties of our calculations are discussed in subsection~\ref{subsec:systematics}, where we also present an extrapolation of the thermodynamic observables to the large-$N$ limit.

\subsection{Results from the simulation ensemble A}
\label{subsec:Nt5_results}

\begin{figure}
\centerline{\includegraphics[width=.80\textwidth]{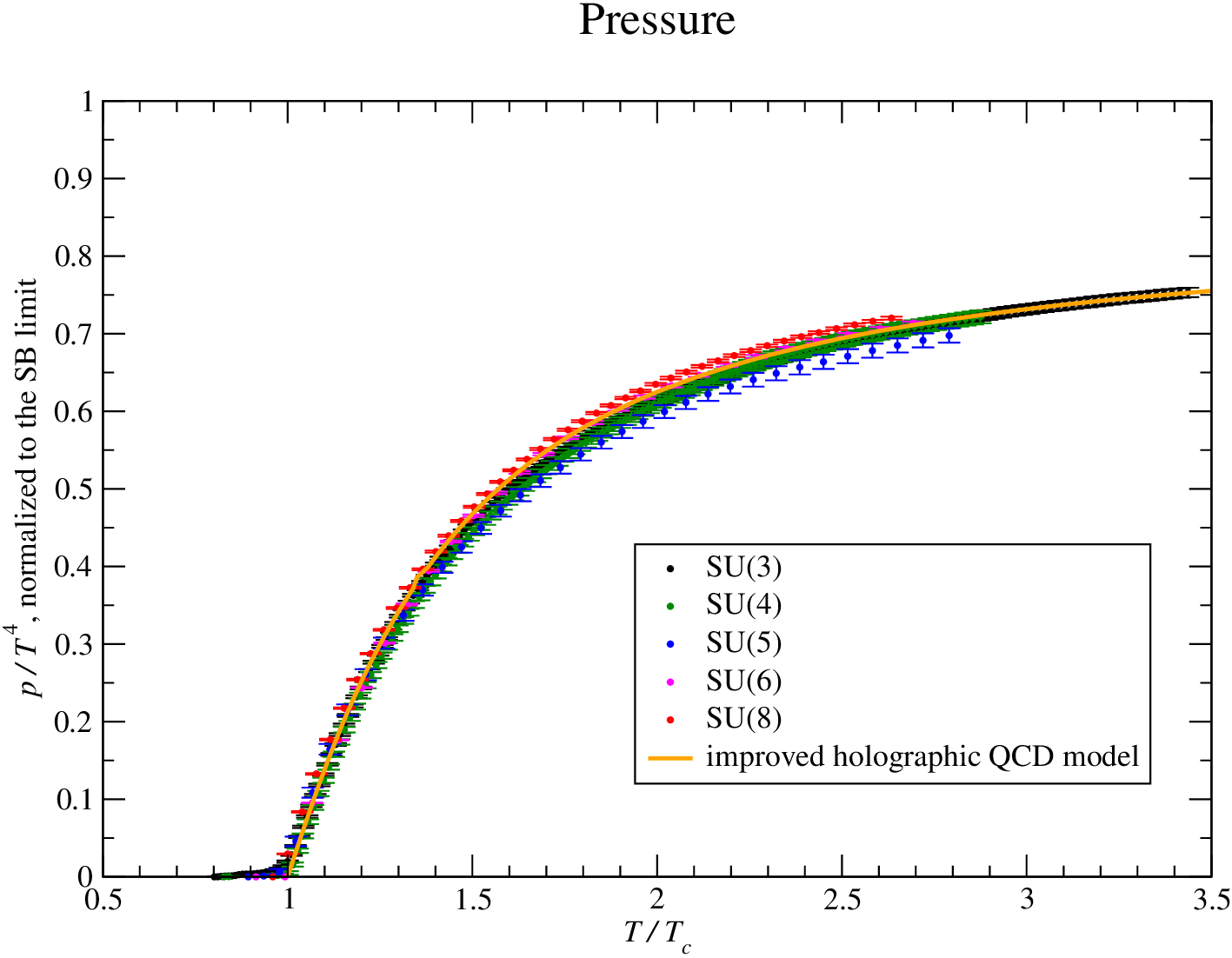}}
\caption{(Color online) The dimensionless ratio $p/T^4$, normalized to the lattice SB limit $\pi^2(N^2-1)R_I(N_t)/45$, \emph{versus} $T/T_c$, as obtained from simulations of $\SU(N)$ lattice gauge theories on $N_t=5$ lattices. Errorbars denote statistical uncertainties only. The results corresponding to different gauge groups are denoted by different colors, according to the legend. The yellow solid line denotes the prediction from the improved holographic QCD model from ref.~\protect\cite{Gursoy:2009jd} (with a trivial, parameter-free rescaling to our normalization).}
  \label{fig:pressure}
\end{figure}

\begin{figure}
\centerline{\includegraphics[width=.80\textwidth]{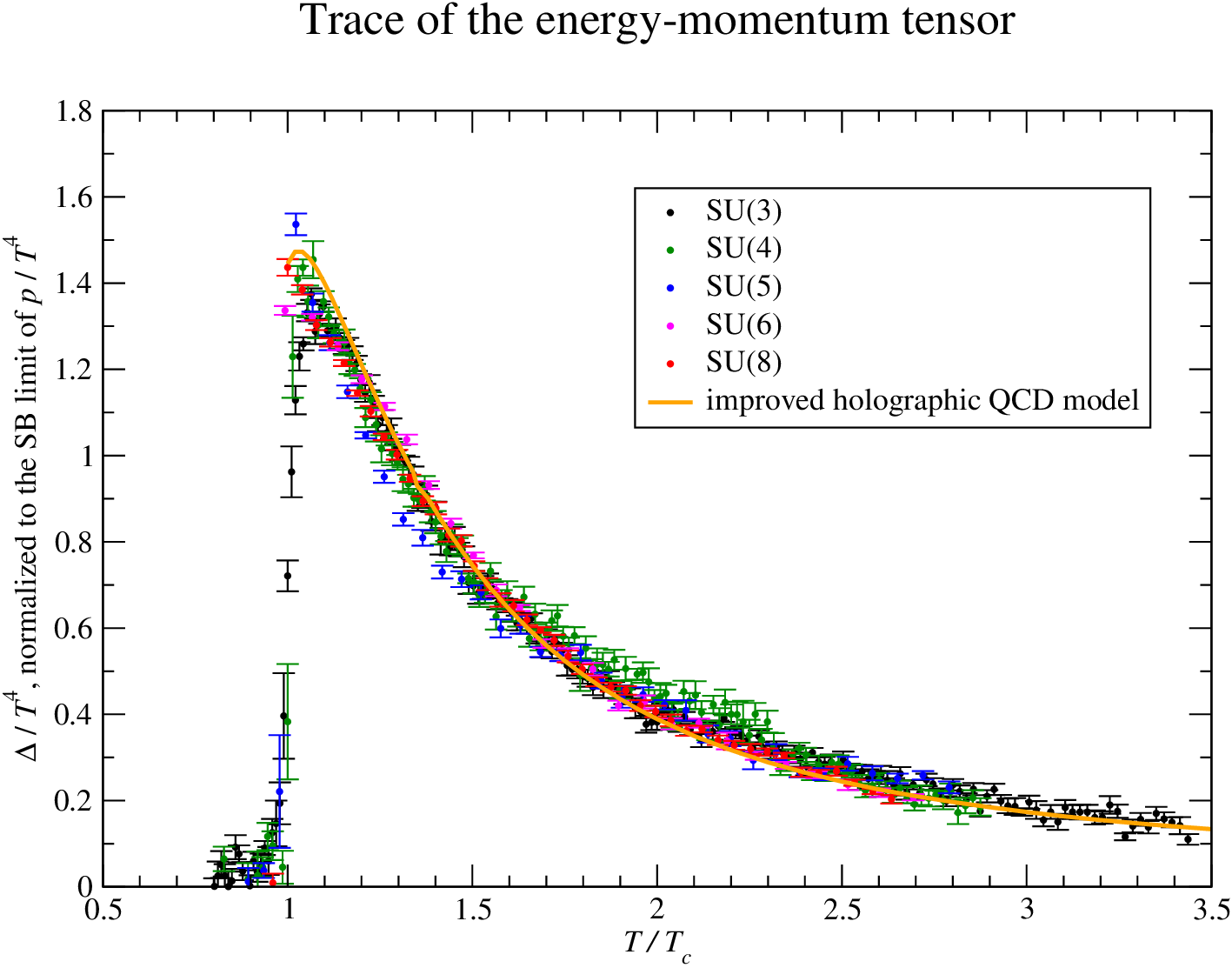}}
\caption{(Color online) Same as in fig.~\ref{fig:pressure}, but for the $\Delta/T^4$ ratio, normalized to the SB limit of $p/T^4$.}
  \label{fig:rescaled_trace}
\end{figure}

\begin{figure}
\centerline{\includegraphics[width=.80\textwidth]{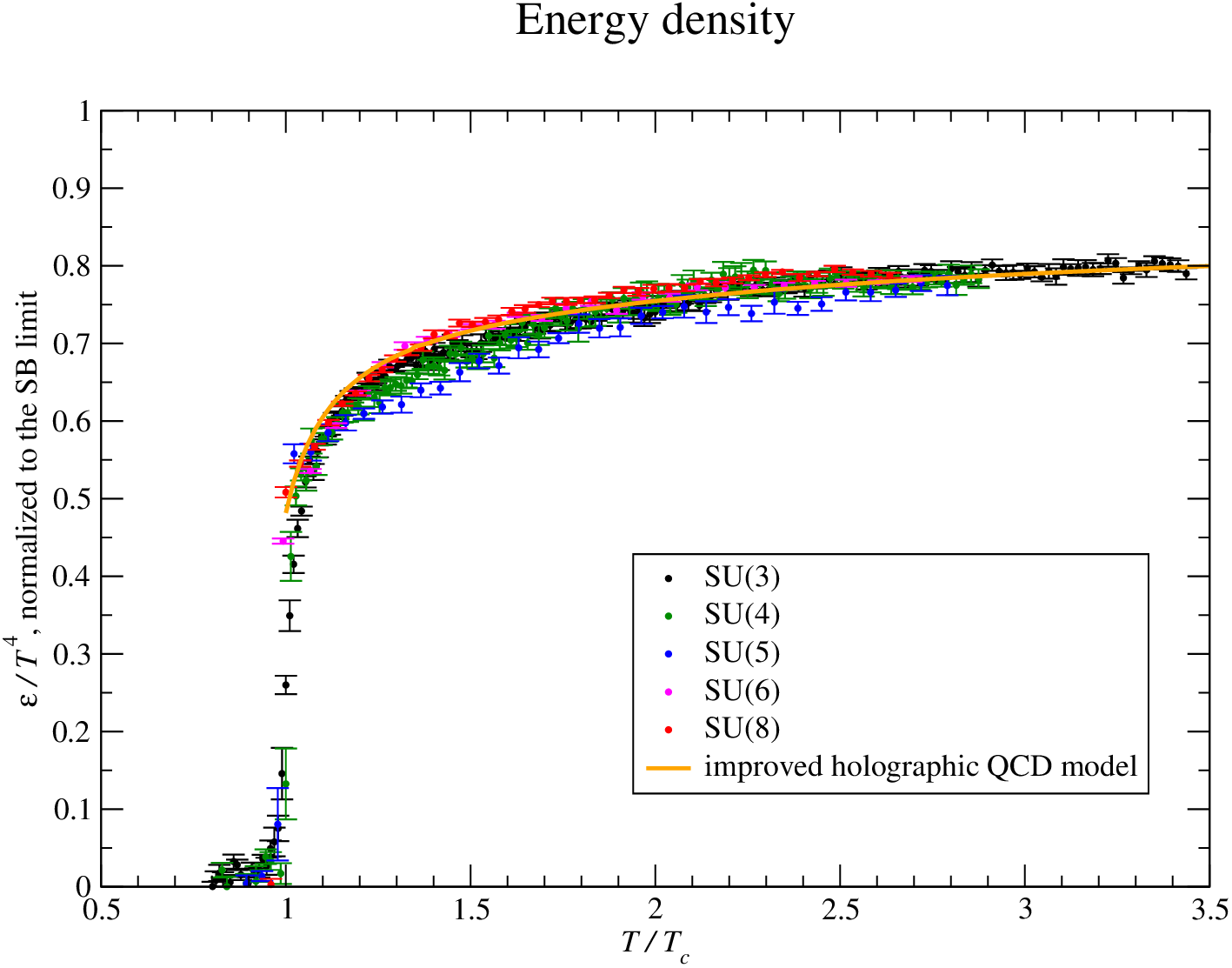}}
\caption{(Color online) Same as in fig.~\ref{fig:pressure}, but for the $\epsilon/T^4$ ratio, normalized to the SB limit.}
  \label{fig:energy}
\end{figure}

\begin{figure}
\centerline{\includegraphics[width=.80\textwidth]{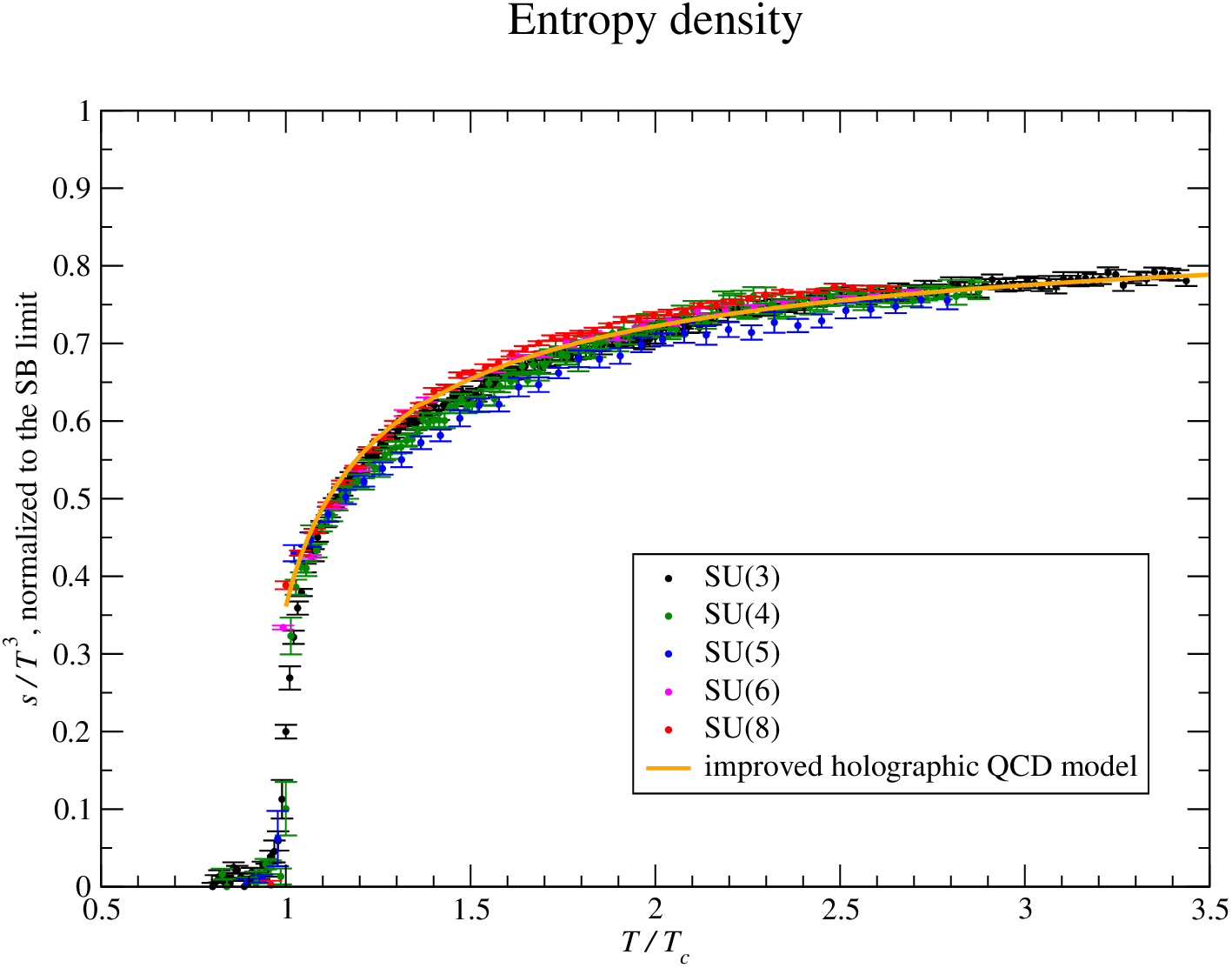}}
\caption{(Color online) Same as in fig.~\ref{fig:pressure}, but for the $s/T^3$ ratio, normalized to the SB limit.}
  \label{fig:entropy}
\end{figure}

The data obtained from our high-precision simulations reveal very strong qualitative and quantitative similarities between the results (normalized to their lattice SB limits) for the thermodynamic observables, in all of the gauge groups investigated,\footnote{This feature has an interpretation in the quasi-particle model discussed in refs.~\cite{Peshier:1995ty, Peshier:1999ww}.} in agreement with the findings of refs.~\cite{Bringoltz:2005rr, Datta:2009tj}. This is manifest in figs.~\ref{fig:pressure}, \ref{fig:rescaled_trace}, \ref{fig:energy} and \ref{fig:entropy}, where we show the dimensionless ratios $p/T^4$, $\Delta/T^4$, $\epsilon/T^4$ and $s/T^3$, normalized to their respective SB limits,\footnote{Since the SB limit of $\Delta$ is zero, we conventionally normalize $\Delta/T^4$ to the SB limit of $p/T^4$.} are plotted against $T/T_c$. 

In these plots, we also show the comparison with the curves obtained from the improved holographic QCD model in ref.~\cite{Gursoy:2009jd}, assuming eq.~(\ref{IHQCD_potential}) as an \emph{Ansatz} for the dilaton potential, and by fitting two free parameters via a comparison with lattice results taken from other works. 

Note that in ref.~\cite{Gursoy:2009jd}, by looking at the comparison with the $N=3$ lattice results for $\Delta/T^4$ from ref.~\cite{Boyd:1996bx}, it was pointed out that the slight discrepancy between the improved holographic QCD model and the lattice results in the region of the peak (which for the $\SU(3)$ lattice data is located at $T \simeq 1.1 T_c$, and is slightly lower than the IHQCD model curve) was likely to be a finite-volume lattice artifact. Here, fig.~\ref{fig:rescaled_trace} indeed confirms this: our results for the $\SU(3)$ gauge group are consistent with ref.~\cite{Boyd:1996bx}, while the $\Delta/T^4$ maximum for the $N>3$ gauge groups is higher and located closer to $T_c$. This is related to the fact that the deconfining phase transition, which is a weakly first-order one for $\SU(3)$, becomes stronger when $N$ is increased~\cite{Lucini:2002ku, Lucini:2003zr, Lucini:2005vg}---see also refs.~\cite{Holland:2003kg, Holland:2003mc, Pepe:2004rc}---and correspondingly the value of the correlation length at the critical point becomes shorter.

\begin{figure}
\centerline{\includegraphics[width=.80\textwidth]{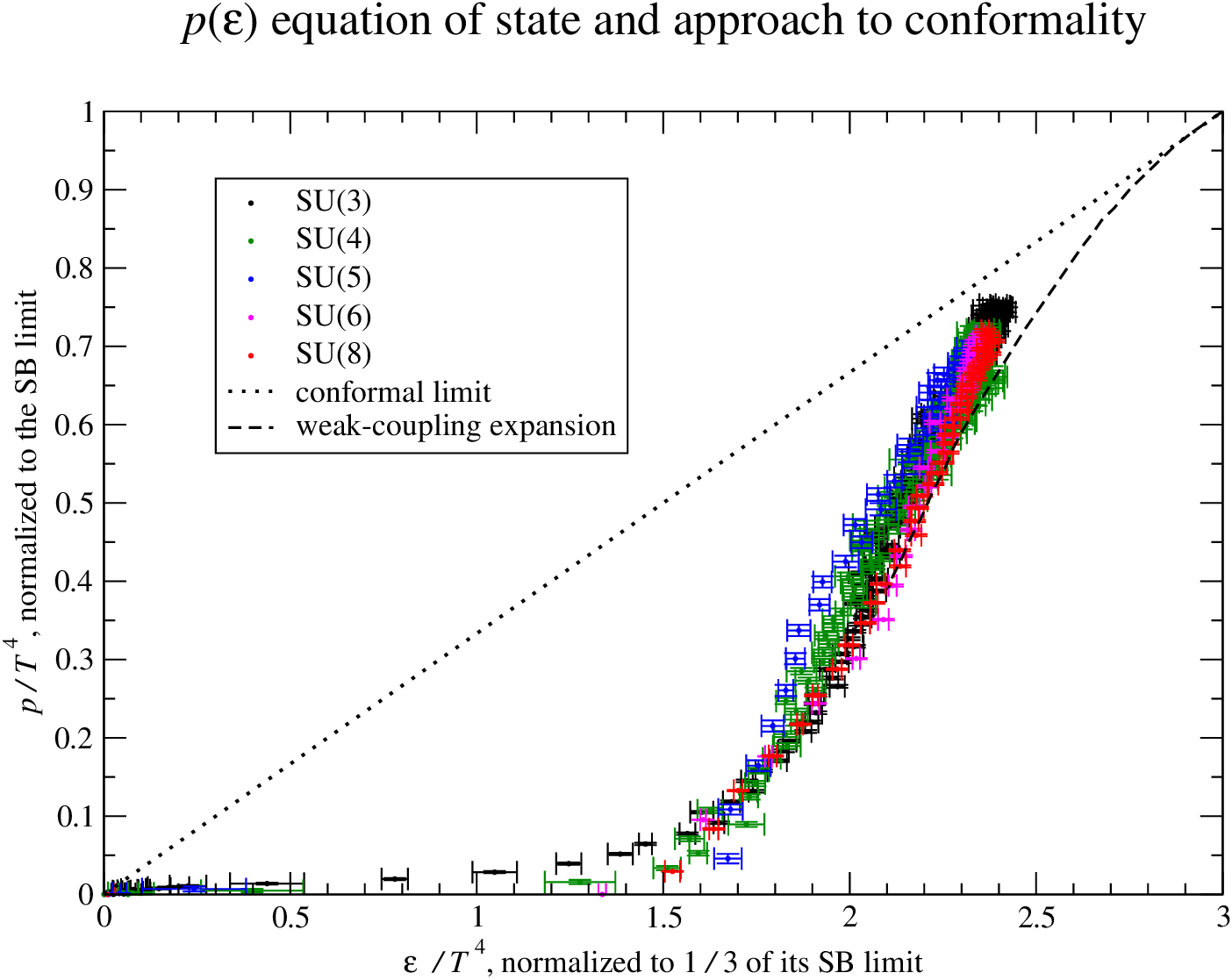}}
\caption{(Color online) The $\SU(N)$ lattice equation of state, expressed as $p(\epsilon)$, and the approach to conformality (dotted line). The dashed curve denotes the prediction from the $\SU(3)$ weak-coupling expansion~\protect\cite{Hietanen:2008tv}.}
  \label{fig:pressure_versus_energy}
\end{figure}

In order to study the relevance of AdS/CFT effective models for the sQGP, it is interesting to investigate `how close' to being conformal the deconfined system is. To address this issue, in fig.~\ref{fig:pressure_versus_energy} we plot the lattice equation of state in a $(\epsilon,p)$ plane (in a format similar to analogous plots in  refs.~\cite{Gavai:2004se, Datta:2009tj}, up to a different normalization of the horizontal axis), where the conformal limit is described by the straight line through the origin and the point corresponding to the SB limit in the top-right corner. As the temperature is increased, the points tend to approach the conformal line from below. As it has been pointed out in ref.~\cite{Datta:2009tj} (by looking at results of $\SU(3)$ and $\SU(4)$ gauge groups), this behavior favors perturbation theory~\cite{Hietanen:2008tv} over a conformal description; our data, displayed in the same format as in ref.~\cite{Datta:2009tj}, appear to confirm this result. However, at temperatures about $3T_c$, although both $\epsilon$ and $p$ are still far from their SB values, the points (for all groups) lie close to the conformal line, so it is reasonable to expect that an AdS/CFT model could also provide a good effective description at these temperatures. 

In this respect, one observable which is particularly interesting to consider is the entropy density $s$, as in the literature it is often mentioned that the relative entropy deficit (with respect to the non-interacting limit) that is observed in lattice QCD simulations is close to the one predicted by AdS/CFT for the strongly-coupled limit of $\mathcal{N}=4$ SYM, which is $1/4$~\cite{Gubser:1996de}. In fact, AdS/CFT also predicts the leading dependence of the $s/s_0$ ratio (with $s_0$ denoting the entropy density in the free case) on the 't~Hooft coupling $\lambda$ of the supersymmetric theory, at the first perturbative order around the strong-coupling limit~\cite{Gubser:1998nz}. However, this information cannot be directly used for a comparison with the YM model, because the 't~Hooft coupling of the supersymmetric model cannot be identified with its counterpart in the non-supersymmetric gauge model: the former does not run with the energy and is simply a parameter of the theory, while the latter is a renormalized coupling which runs with the momentum scale and, in general, is scheme-dependent. One possible strategy to bypass this potential ambiguity and try to work out a comparison between the supersymmetric model and the YM theory could be by considering the screening masses~\cite{Bak:2007fk}. 

In general, insisting on identifying the $\lambda$ parameter of $\mathcal{N}=4$ SYM with some particular definition of the running 't~Hooft coupling in the YM theory would be a scheme-dependent, and thus ambiguous, procedure. However, one may still argue that one could get at least some general, semi-quantitative information about how well the supersymmetric model captures the behavior of the YM theory (in the regime where it becomes `quasi-conformal') by identifying the supersymmetric theory parameter with the YM running coupling evaluated in some physically-motivated, `good' scheme (meaning a scheme for which low-order perturbation theory calculations work well down to energies close to the non-perturbative scale of the theory). 

\begin{figure}
\centerline{\includegraphics[width=.80\textwidth]{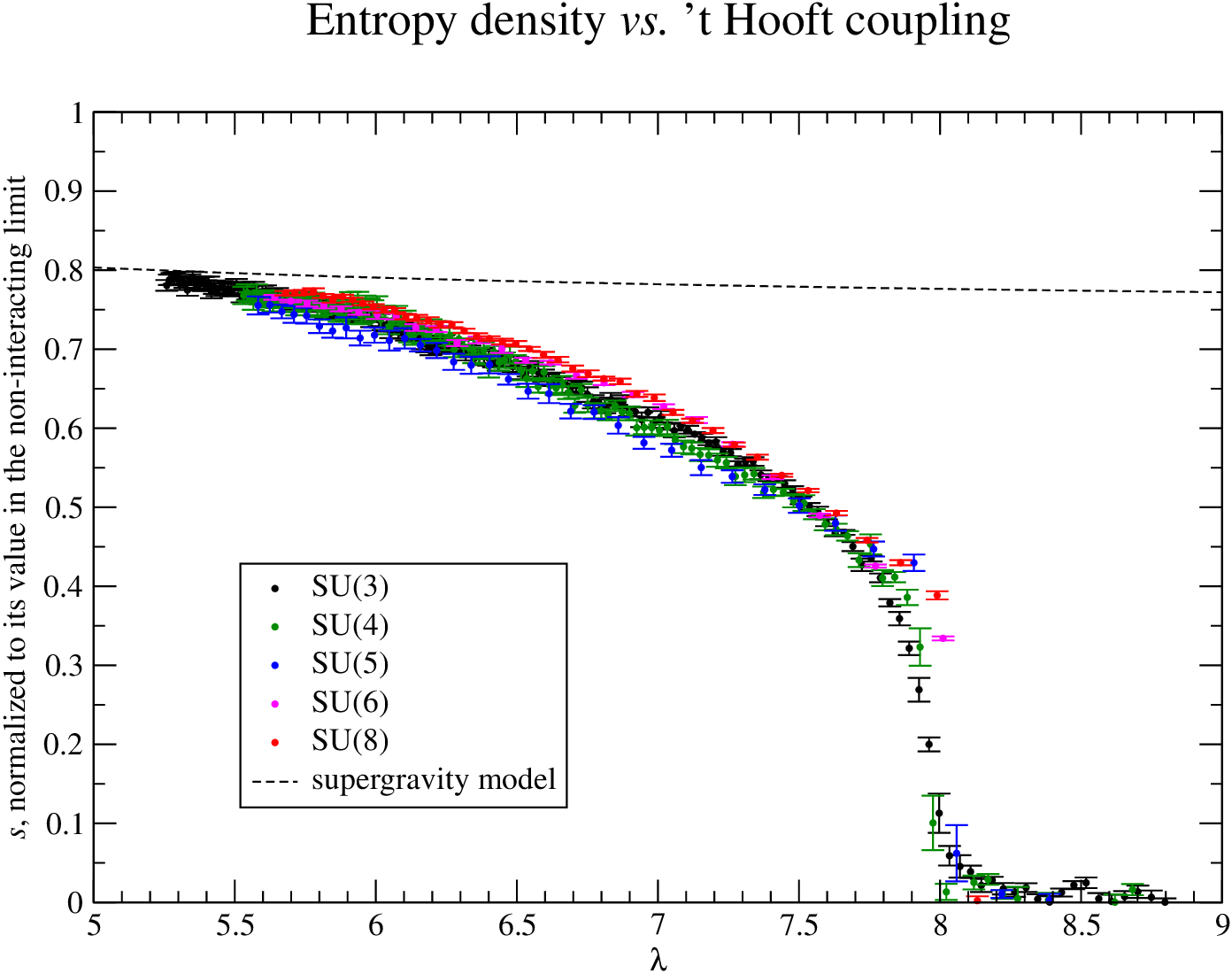}}
\caption{(Color online) The entropy density $s$ (normalized to its limit for the non-interacting gluon gas $s_0$) \emph{versus} the renormalized 't~Hooft coupling $\lambda$ evaluated in the $\overline{\rm MS}$ scheme at the momentum scale characteristic of the corresponding temperature, as estimated from two-loop perturbation theory; note that the data correspond to temperatures which are increasing from left to right. The dashed curve is the AdS/CFT prediction for the dependence of $s/s_0$ on the 't~Hooft coupling, at the leading order in an expansion around the strong-coupling limit, eq.~(\protect\ref{sugra_entropy}): intriguingly, if one identifies the 't~Hooft coupling of the supersymmetric model with the running coupling $\lambda$ of the YM theory in the $\overline{\rm MS}$ scheme, the numerical results in the temperature regime where the deconfined plasma approaches a quasi-conformal regime---see fig.~\protect\ref{fig:pressure_versus_energy}---tend to get very close to the curve describing the AdS/CFT model. However, see the text for a discussion of the ambiguity in such a comparison.}
\label{fig:entropy_lambda}
\end{figure}

In this context, we mention that in the recent literature there are various studies discussing the running of the coupling in large-$N$ lattice gauge theories, with different approaches and in different schemes~\cite{deForcrand:2005pb, deForcrand:2005rg, Allton:2008ty, Lucini:2008vi}, pointing to a mild, well-behaved dependence on $N$. In particular, one popular choice to define the running coupling is the Schr\"odinger functional (SF) scheme~\cite{Luscher:1991wu, Luscher:1992an}; the behavior of the running coupling in the SF scheme in the $\SU(4)$ lattice gauge theory has been investigated by Lucini and Moraitis~\cite{Lucini:2008vi}: they found good agreement with two-loop perturbation theory down to energies comparable to the square root of the string tension. Assuming that such agreement also holds in the large-$N$ limit~\cite{Allton:2008ty, Lucini:2008vi}, one can estimate the running coupling for all gauge groups considered in the present work, and convert between different schemes---including, in particular, the modified minimal-subtraction ($\overline{\rm MS}$) scheme---using the expressions for $\Lambdamsbar/\sqrt{\sigma}$ and $T_c/\sqrt{\sigma}$ at different $N$ which can be taken from the literature~\cite{Allton:2008ty, Lucini:2005vg}. Following this approach, one can obtain a plot of the entropy \emph{versus} an approximate estimate of the running 't~Hooft coupling in the $\overline{\rm MS}$ scheme, which is shown in fig.~\ref{fig:entropy_lambda}; at temperatures around $3T_c$ (where the plasma approaches a `quasi-conformal' regime), the physical 't~Hooft coupling for $\SU(3)$ is around $5.5$ (in agreement with ref.~\cite{Boyd:1996bx}), and still large for all $\SU(N)$ groups. 

Incidentally, it may be interesting to note that, if one were to directly identify the $\lambda$ parameter of the $\mathcal{N}=4$ model with the YM running 't~Hooft coupling evaluated in the $\overline{\rm MS}$ scheme, then eq.~(\ref{sugra_entropy}) would predict a curve (dashed line in fig.~\ref{fig:entropy_lambda}) that is smoothly approached by the lattice results in the temperature regime where the AdS/CFT description is expected to begin to work (right end of the plot). Also, the values of the 't~Hooft coupling in that regime agree with the estimate obtained in ref.~\cite{Gubser:2006qh} from the comparison of the drag force on heavy quarks in the $\mathcal{N}=4$ model and in lattice QCD simulations with dynamical quarks, and appear to be roughly comparable with the lower bound of the window for this parameter identified in ref.~\cite{Noronha:2009vz}, namely: $\lambda \approx 10$--$25$ (although the gravity model discussed there also includes a Gauss-Bonnet term~\cite{Buchel:2008vz}, and the constraints on the parameters are derived by comparing with experimental data and with results of lattice QCD simulations with dynamical fermions~\cite{Aoki:2005vt, Bernard:2006nj, Cheng:2007jq, Bazavov:2009zn}). 

However, we stress once more that the comparison suggested by fig.~\ref{fig:entropy_lambda} should be taken \emph{cum grano salis}. First and foremost, because of the ambiguity and scheme-dependence in comparing a running coupling of the non-supersymmetric model with a parameter of the conformal, supersymmetric model, as we discussed above. Secondly, because, even if one could prove that the identification of the $\lambda$ parameter of the SYM theory with the $\overline{\rm MS}$ 't~Hooft coupling is well-motivated and should be preferred over other schemes, it is not at all obvious that, at the (relatively small) $\lambda$-values where the YM plasma appears to become quasi-conformal, subleading corrections not included in eq.~(\ref{sugra_entropy}) would still be negligible. Finally, because it should be noted that the $\lambda$-dependence predicted by eq.~(\ref{sugra_entropy}) in a range of couplings comparable to those shown in fig.~\ref{fig:entropy_lambda} is relatively weak, therefore it may well be that the identification of the SYM $\lambda$ parameter with the YM running coupling evaluated in some different scheme would yield equally good optical agreement; in any case, the systematic and statistical uncertainties involved do not allow one to draw any firm conclusion about a ``preferred'' scheme for this type of comparison.

Another interesting issue concerns the possibility of non-perturbative $T^2$ contributions to $\Delta$, which may play a prominent r\^ole in the temperature range between the maximum of $\Delta/T^4$ (which is attained at $1.1 T_c$ for $\SU(3)$, and tends to values closer to $T_c$ when $N$ is increased) and the onset of the regime where perturbation theory is accurate: as it has been pointed out by Pisarski~\cite{Pisarski:2006hz, Pisarski:2006yk}, the $\SU(3)$ data in this regime~\cite{Boyd:1996bx} appear to be described well by eq.~(\ref{Pisarski_pressure}).

\begin{figure}
\centerline{\includegraphics[width=.80\textwidth]{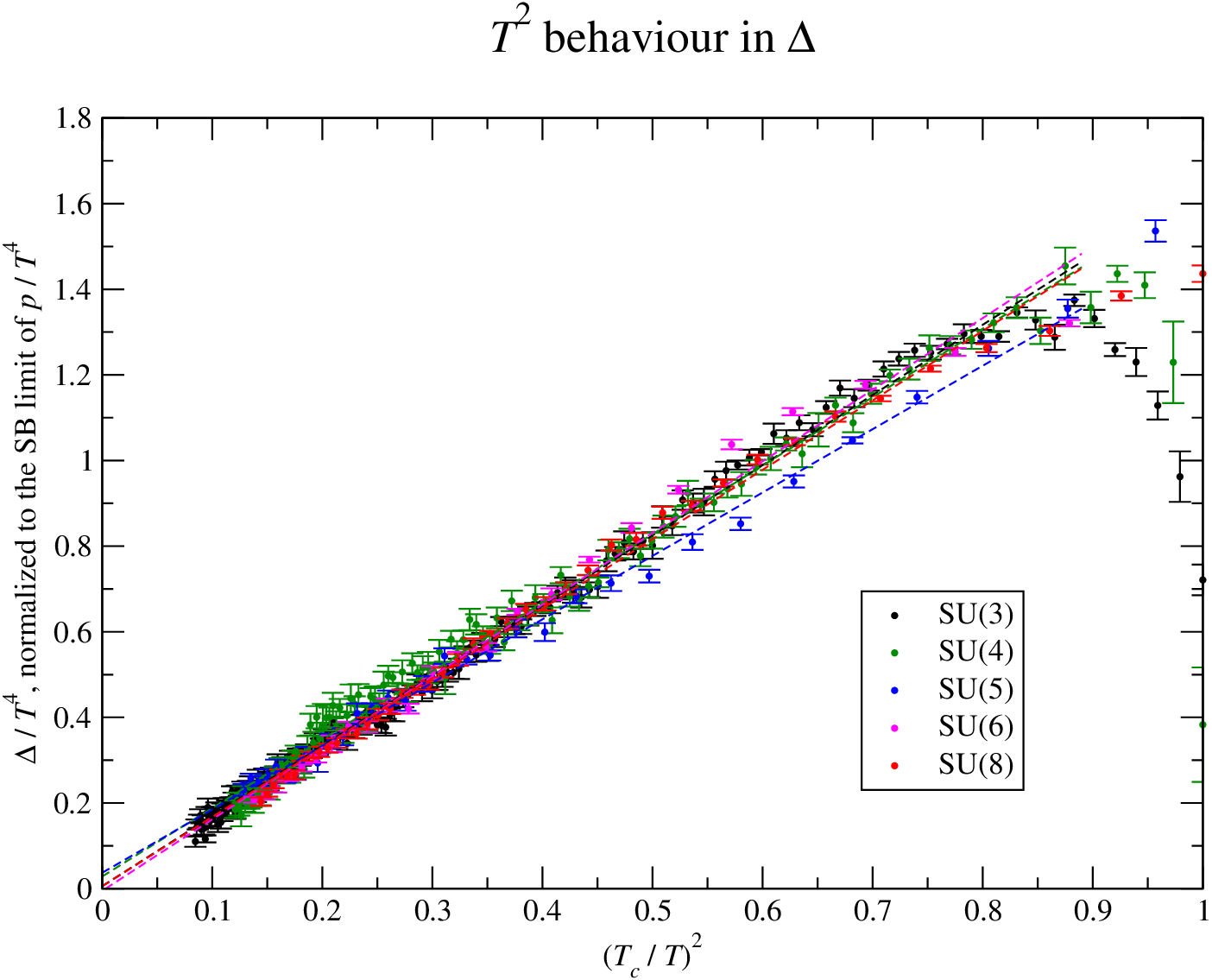}}
\caption{(Color online) Does $\Delta$ receive non-perturbative $\mathcal{O}(T^2)$ contributions in the range between the deconfinement temperature $T_c$ and the onset of the perturbative regime? Do such contributions appear only in $\SU(3)$, or are they a feature of the large-$N$ limit too? This diagram shows that the data of fig.~\protect\ref{fig:rescaled_trace} for the $\Delta/T^4$ ratio, plotted against $(T_c/T)^2$, may be compatible with a linear behavior. The dashed straight lines are obtained fitting the numerical results to eq.~(\protect\ref{Pisarski_linear_fit}) in the range $(T_c/T)^2 \le 0.9$, with $m=1.64(1)$, $1.60(2)$, $1.480(9)$, $1.67(1)$, $1.622(9)$ and $q=0.006(3)$, $0.028(6)$, $0.037(5)$, $-0.005(3)$, $0.005(4)$ for $\SU(3)$, $\SU(4)$, $\SU(5)$, $\SU(6)$ and $\SU(8)$, respectively.}
  \label{fig:Pisarski}
\end{figure}

Our data seem to confirm that this a general feature of all the gauge groups that we studied in this work: fig.~\ref{fig:Pisarski} shows that the results for $\Delta/T^4$ (which are plotted with the same normalization as in fig.~\ref{fig:rescaled_trace}) can be described by:
\eq{Pisarski_linear_fit}
\Delta/T^4=\frac{\pi^2}{45}(N^2-1) R_I(N_t) \cdot \left[ m \left( \frac{T_c}{T}\right)^2 +q \right].
\en
In ref.~\cite{Megias:2009mp}, it was argued that this behavior may be related with the existence of a dimension-two condensate above the phase transition.

In particular, the parameters obtained from linear fits according to eq.~(\ref{Pisarski_linear_fit}) show a weak $N$-dependence. However, the precision of our results does not allow us to rule out the possibility that the data may actually be described by different, more complicated functional forms (possibly involving combinations of logarithms of perturbative origin).

\subsection{Systematic uncertainties, extrapolation to the continuum and to the large-$N$ limit}
\label{subsec:systematics}

Having presented the numerical results from our main ensembles of configurations (which have been obtained on lattices with $N_t=5$ for the finite-temperature runs), we now turn to a discussion of the main systematic errors in our calculations, and to the issues related to the continuum and large-$N$ extrapolations.

The main systematic uncertainties involved in our calculation include the setting of the physical scale, discretization effects, and finite-volume effects.

To set the physical scale for our simulations, we used high-precision non-perturbative results which are available in the literature. In particular, for the $\SU(3)$ gauge group, we set the scale using the $r_0$ parameter, via the interpolating formula given by eq.~(2.6) in ref.~\cite{Necco:2001xg}, which holds for $5.7 \le \beta \le 6.92$, and thus completely covers the range of points that we simulated. Therefore,  the systematic uncertainty associated with the scale determination for our $\SU(3)$ data receives contributions from the accuracy limits on that interpolating formula (which are between $0.5\%$ and $1\%$), and from the ambiguity in choosing the dimensionful quantity through which the scale is set: different observables would lead to slight, $\mathcal{O}(a^2)$, differences in the scale determination. For the $\SU(N>3)$ groups, we set the scale by interpolating the string tension values calculated by Lucini, Teper and Wenger~\cite{Lucini:2004my, Lucini:2000qp}; since these results do not extend up to the largest $\beta$-values that we simulated, in the latter range we determined the scale following the strategy discussed in ref.~\cite{Allton:2008ty}, namely through the three-loop perturbative lattice $\beta$-function~\cite{Alles:1996cy} in the mean-field improved lattice scheme~\cite{Parisi:1980pe}, taking lattice corrections into account (note that in ref.~\cite{Allton:2008ty} it was shown that the two methods are, in fact, consistent). The systematic uncertainty associated with the scale determination is difficult to quantify, but can be (roughly) estimated by comparing the results obtained using different interpolations for $\log(\sigma a^2)$ as a function of $(\beta-\beta_\star)$ (where $\beta_\star$ is a value chosen within the interpolating range). We estimate the systematic uncertainty related to the scale determination to be up to $2\%$.

To get a reliable extrapolation to the continuum limit, in principle one should repeat the whole calculation on a series of lattices of increasingly finer spacing, \emph{i.e.} at increasingly larger $N_t$ values (keeping the $N_t/N_s$ ratio fixed, to disentangle from possible finite-volume effects), and perform a linear extrapolation in $a^2$ to the limit $a \to 0$ for each gauge group separately. Unfortunately, for a set of high-precision calculations like the present one, this task would be computationally very demanding. On the other hand, in the literature there are indications suggesting that the results from our simulation ensemble A, obtained on lattices with $N_t=5$, may already be relatively close to the continuum limit. In particular, ref.~\cite{Datta:2009tj} discusses a recent calculation similar to ours, finding that, for $\SU(4)$, the continuum limit is reached at $N_t=6$. 

\begin{figure}
\centerline{\includegraphics[width=.80\textwidth]{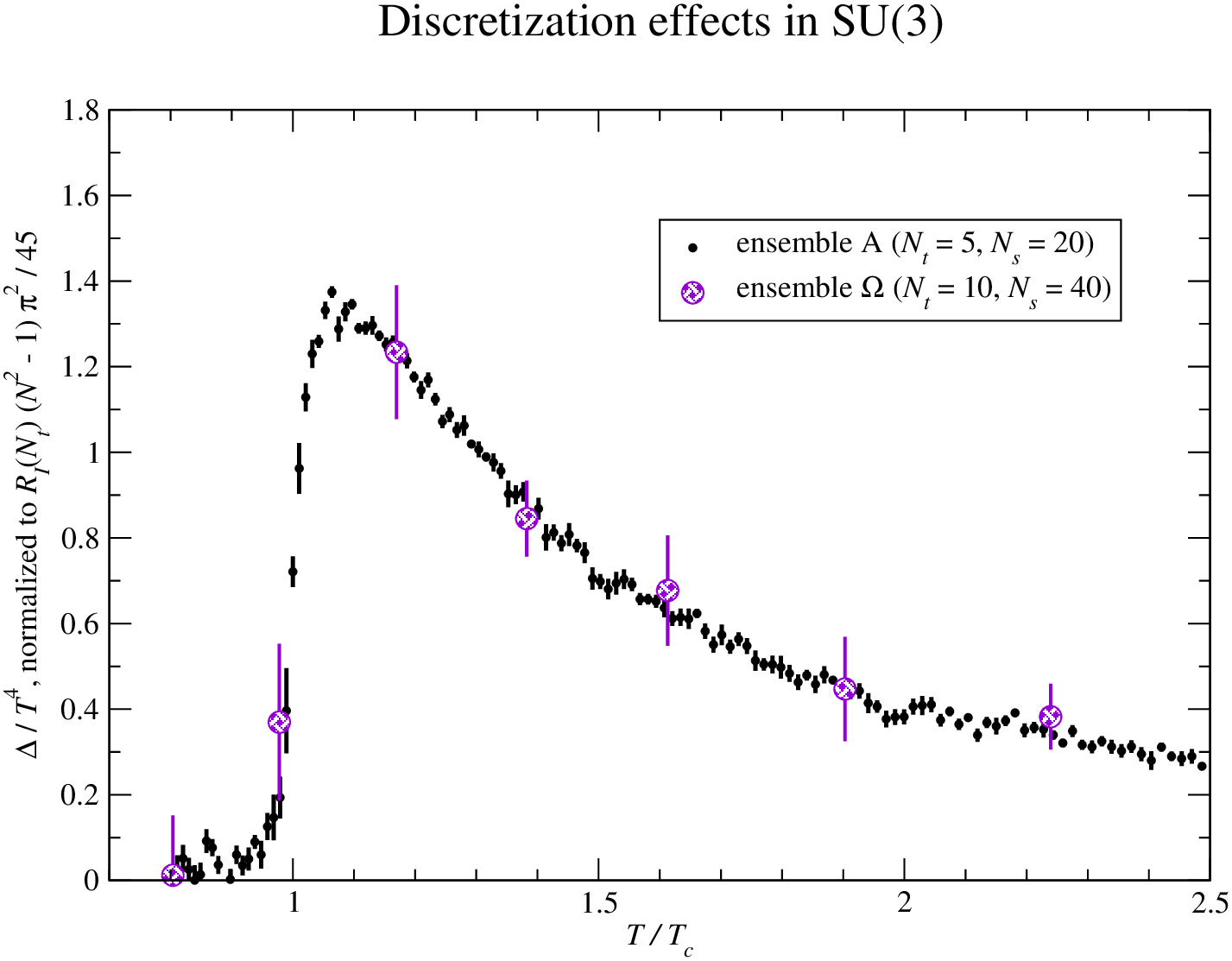}}
\caption{(Color online) Comparison of the results for the trace of the energy-momentum tensor in $\SU(3)$ obtained from our main set of simulations (ensemble A, corresponding to lattices with $N_t=5$ and $N_s=20$, denoted by the small black symbols) and from a set characterized by a lattice spacing twice as fine (ensemble ${\rm \Omega}$, with $N_t=10$ and $N_s=40$, denoted by the large violet symbols).}
  \label{fig:continuum_extrapolation}
\end{figure}

To check the impact of discretization effects on our results from the $N_t=5$ lattices, we also performed a preliminary set of $\SU(3)$ simulations on lattices with $N_t=10$, $N_s=40$, that we denote as `ensemble ${\rm \Omega}$' in the following---see tab.~\ref{tab:ensemble_Omega_simulation_info} for technical information. However, the comparison between the results for the trace of the energy-momentum tensor in $\SU(3)$, as evaluated from the two ensembles, which is shown in fig.~\ref{fig:continuum_extrapolation}, does not give conclusive information on discretization effects. By comparing with similar works in the literature~\cite{Datta:2009tj}, we take $3\%$ as a rough estimate of the impact of discretization effects (not accounted for by the rescaling of the SB limit on the lattice through the factor $R_I(N_t)$) affecting our main set of results from the ensemble A.

\begin{table}[h]
\centering
\begin{tabular}{|c|cc|cc|cc|}  
\hline
$N$ & $N_s$ & $N_t$ & $T=0$ statistics & finite-$T$ statistics & $\beta$ & $T$ \\
\hline \hline
$3$ & $40$ & $10$ & $9000$ & $2800$ & $6.055$ & $0.804$ \\
 & & & & & $ 6.180 $ & $0.978$ \\
 & & & & & $ 6.305 $ & $1.170$ \\
 & & & & & $ 6.430 $ & $1.383$ \\
 & & & & & $ 6.550 $ & $1.613$ \\
 & & & & & $ 6.680 $ & $1.903$ \\
 & & & & & $ 6.805 $ & $2.239$ \\
\hline
\end{tabular}
\caption{Parameters of the set of lattice simulations (ensemble ${\rm \Omega}$) used to check the impact of discretization effects in the main set of results presented in this work.}\label{tab:ensemble_Omega_simulation_info}
\end{table}

As it concerns finite-volume effects, at the temperatures investigated in the present work one generally expects them to be strongly suppressed, due to screening. In ref.~\cite{Panero:2008mg}, we checked this explicitly for the setup of the present simulations, and found that further, non-trivial infrared corrections to the lattice equation of state~\cite{Gliozzi:2007jh} are not visible within our data precision at the temperatures and volumes considered in the present work.

To summarize, we estimate that the major sources of systematic uncertainties in our simulations come from the determination of the scale and from discretization effects. In particular, in our plots, the scale uncertainty has an impact on both the temperature axes and on the thermodynamic quantities, which have an explicit scale-dependence through eq.~(\ref{lattice_rescaled_trace}). Combining the errors associated to scale determination and discretization effects in quadrature, the systematic uncertainty in our results turns out to be around $3 \%$.

Finally, we perform an extrapolation of the results from our ensemble A for $\Delta/T^4$ to the $N \to \infty$ limit. According to the method we used in this work, this observable is evaluated from direct measurements of plaquette expectation values, see eq.~(\ref{lattice_rescaled_trace}), whereas all other quantities are obtained indirectly from it, as discussed above.

In order to do the extrapolation to the large-$N$ limit, we first interpolate our results for the trace of the stress-energy tensor for the different gauge groups, according the functional form given in eq.~(C1) of ref.~\cite{Bazavov:2009zn} (redefined according to our normalization conventions):
\eq{interpolating_formula_for_Delta}
\frac{\Delta}{T^4} = \frac{\pi^2}{45} (N^2-1) \cdot \left( 
1 - \frac{1}{\left\{
1 + \exp \left[ \frac{(T/T_c)-f_1}{f_2}
\right] \right\}^2}
\right) \left( f_3\frac{T_c^2}{T^2} + f_4\frac{T_c^4}{T^4}
\right),
\en
which models the region near the critical temperature using a modified hyperbolic tangent and includes $T^{-2}$ and $T^{-4}$ terms to mimic the data behavior at temperatures intermediate between the $\Delta/T^4$ maximum and the onset of the perturbative regime. The fit results are given in tab.~\ref{tab:Delta_fit_results}, where the uncertainties shown do not include systematic errors.

\begin{table}[h]
\centering
\begin{tabular}{|c||c|c|c|c||c|}  
\hline
$N$ & $f_1$ & $f_2$ & $f_3$ & $f_4$ & $\chi^2/$d.o.f. \\
\hline \hline
 $3$ & $1.0317(13)$ & $0.02557(96)$  & $1.686(14)$ & $-0.038(24)$ & $2.30$\\
 $4$ & $1.0154(20)$ & $0.0082(12)$    & $1.763(20)$ & $-0.185(33)$ & $2.10$ \\
 $5$ & $1.0094(98)$ & $0.0056(10)$    & $1.729(83)$ & $-0.205(65)$ & $1.56$ \\
 $6$ & $0.989(37)$   & $0.009(87)$      & $1.803(65)$ & $-0.244(97)$ & $2.77$ \\
 $8$ & $0.9944(46)$ & $0.0058(51)$    & $1.720(26)$ & $-0.165(40)$ & $6.11$ \\
\hline
\end{tabular}
\caption{Results of the fits for $\Delta$ according to eq.~(\protect\ref{interpolating_formula_for_Delta}); errors shown do not include systematic uncertainties.}\label{tab:Delta_fit_results}
\end{table}

The parameters thus obtained are then fitted as functions of $N^{-2}$ including a constant, a linear and a quadratic term:
\eq{large_N_parameter_extrapolation}
f_i (N) = c_i + \frac{l_i}{N^2} + \frac{q_i}{N^4};
\en
we note that, for $f_1$, $f_3$ and $f_4$, the $\mathcal{O}(N^{-4})$ term can be neglected, therefore for these parameters we perform a two-parameter fit, setting $q_i$ to zero (and we include the difference between such linear interpolation and a quadratic interpolation as a contribution to the overall systematic effects). On the contrary, for $f_2$ it is necessary to include the $\mathcal{O}(N^{-4})$ term in 
the fit, otherwise one would find an unphysical, negative result for the extrapolated value $c_2$.

\begin{figure}
\centerline{\includegraphics[width=.45\textwidth]{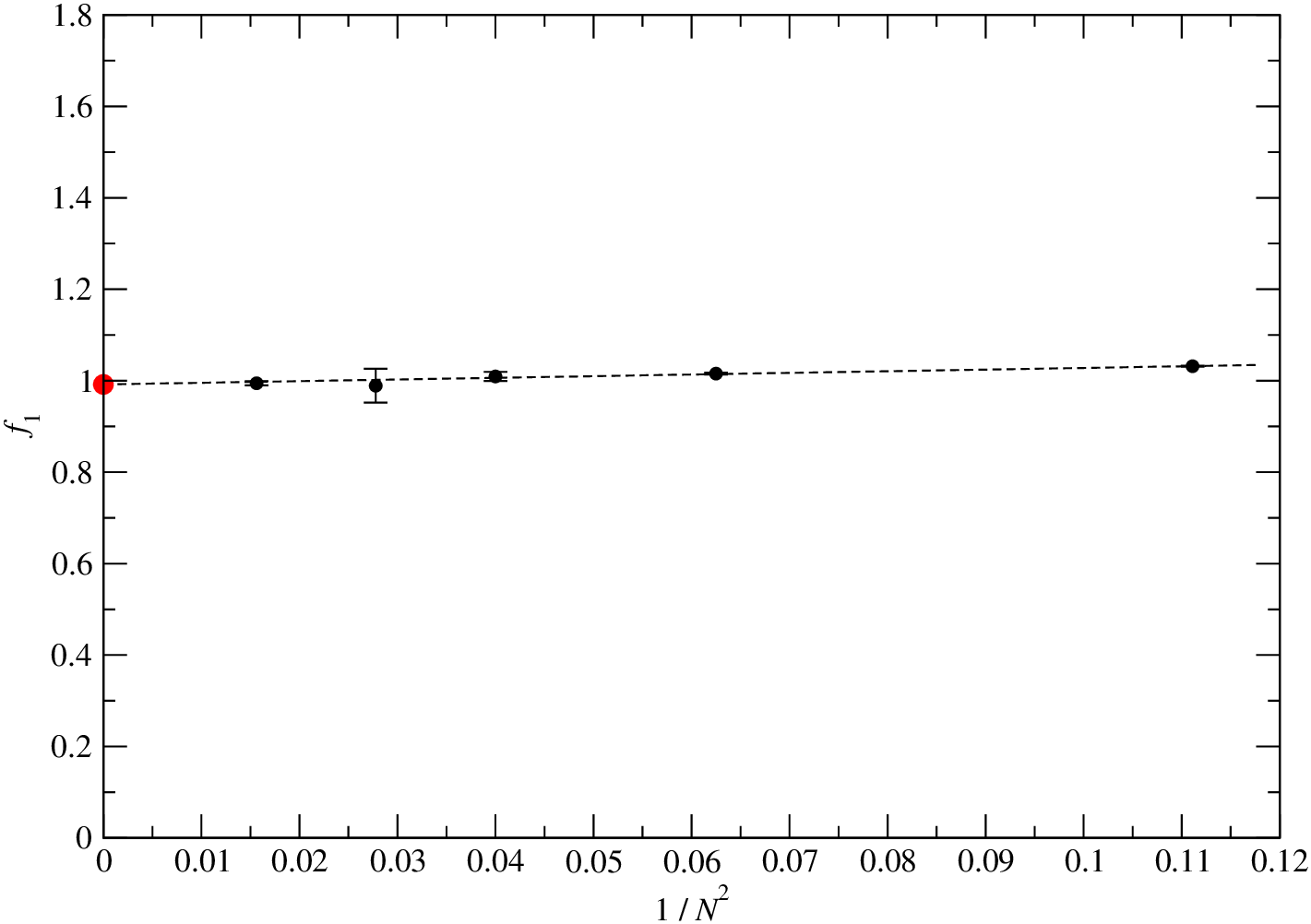}\hspace{1cm}
\includegraphics[width=.46\textwidth]{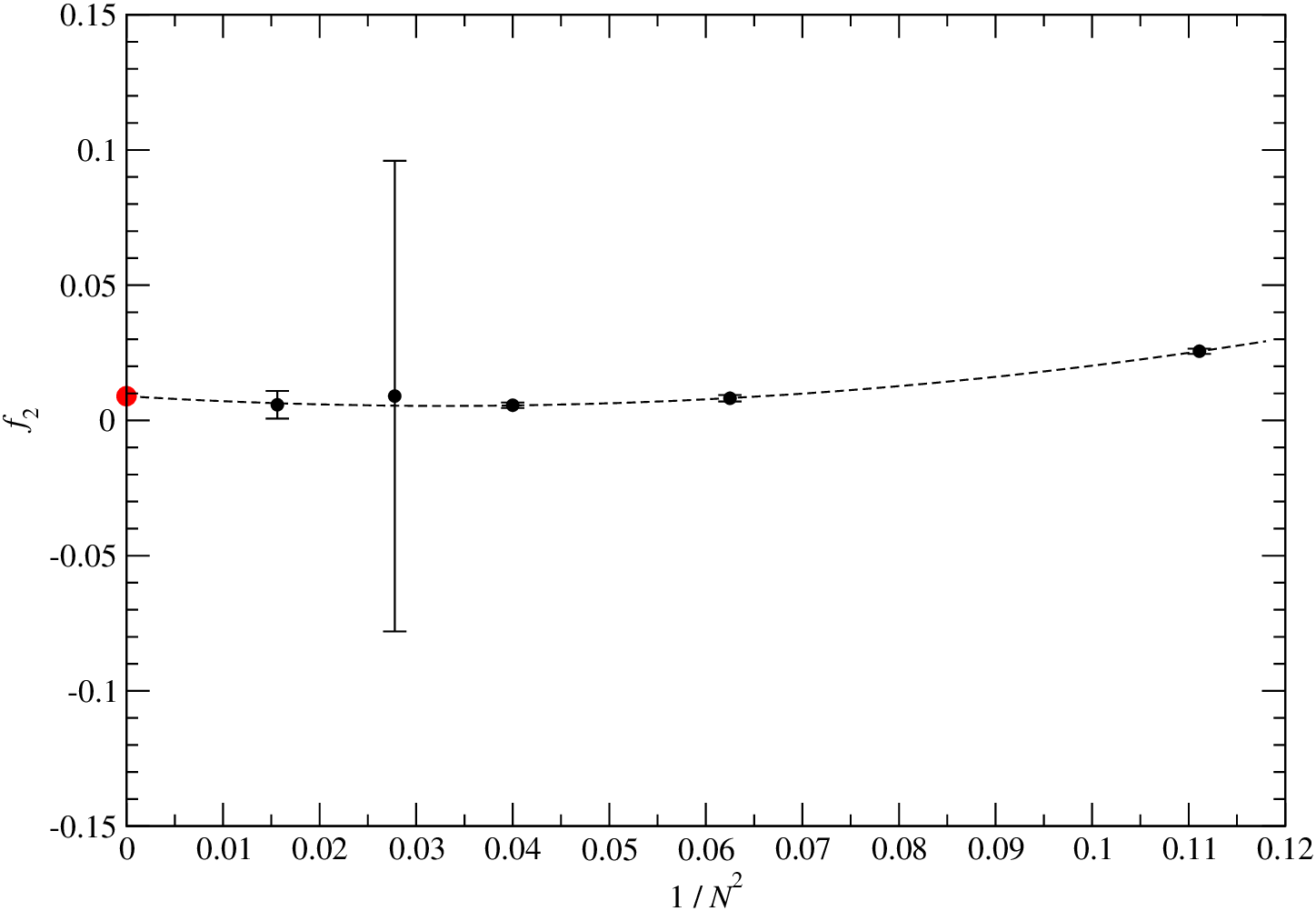}}
\vspace{10mm}
\centerline{\includegraphics[width=.45\textwidth]{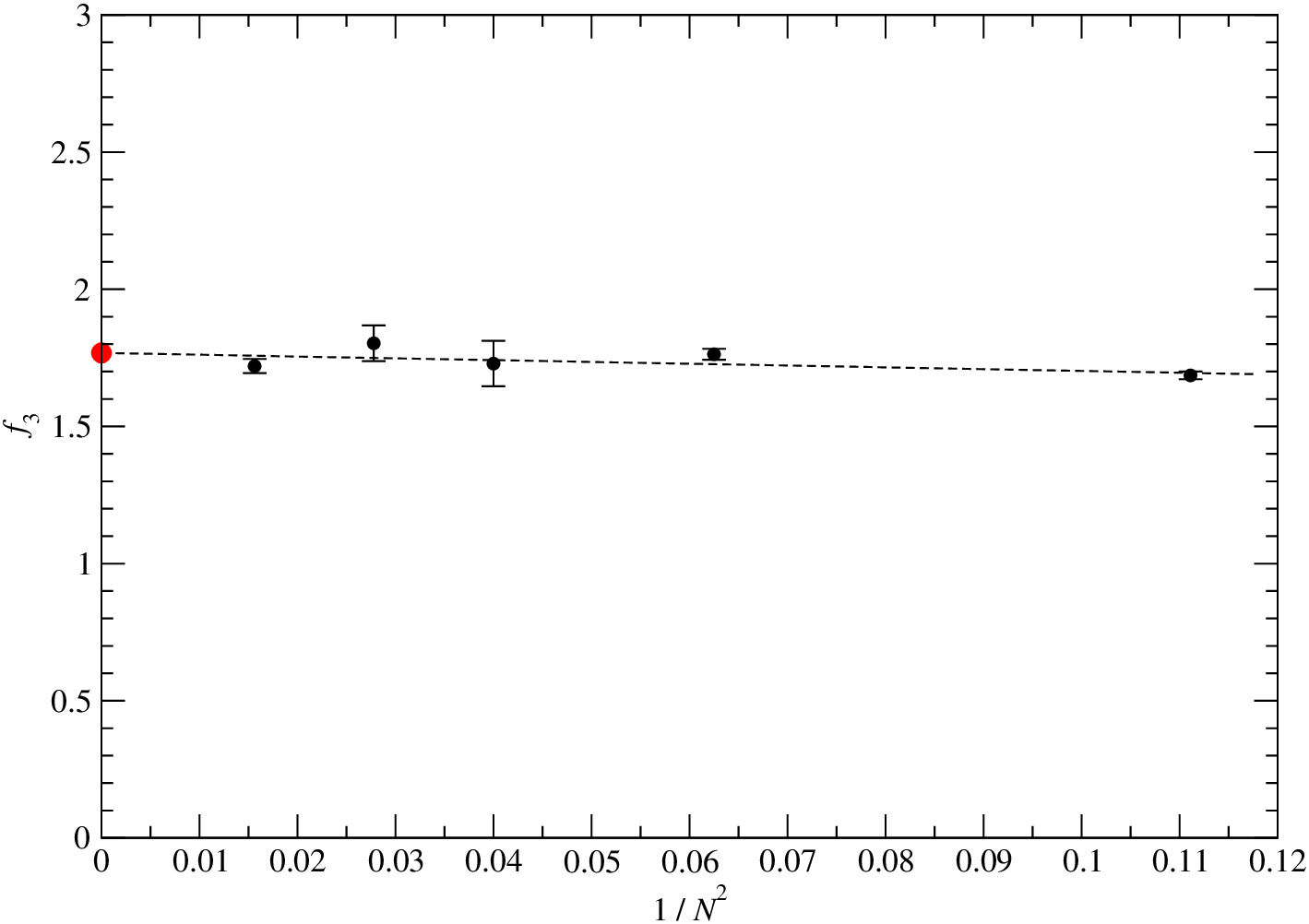}\hspace{1cm}
\includegraphics[width=.46\textwidth]{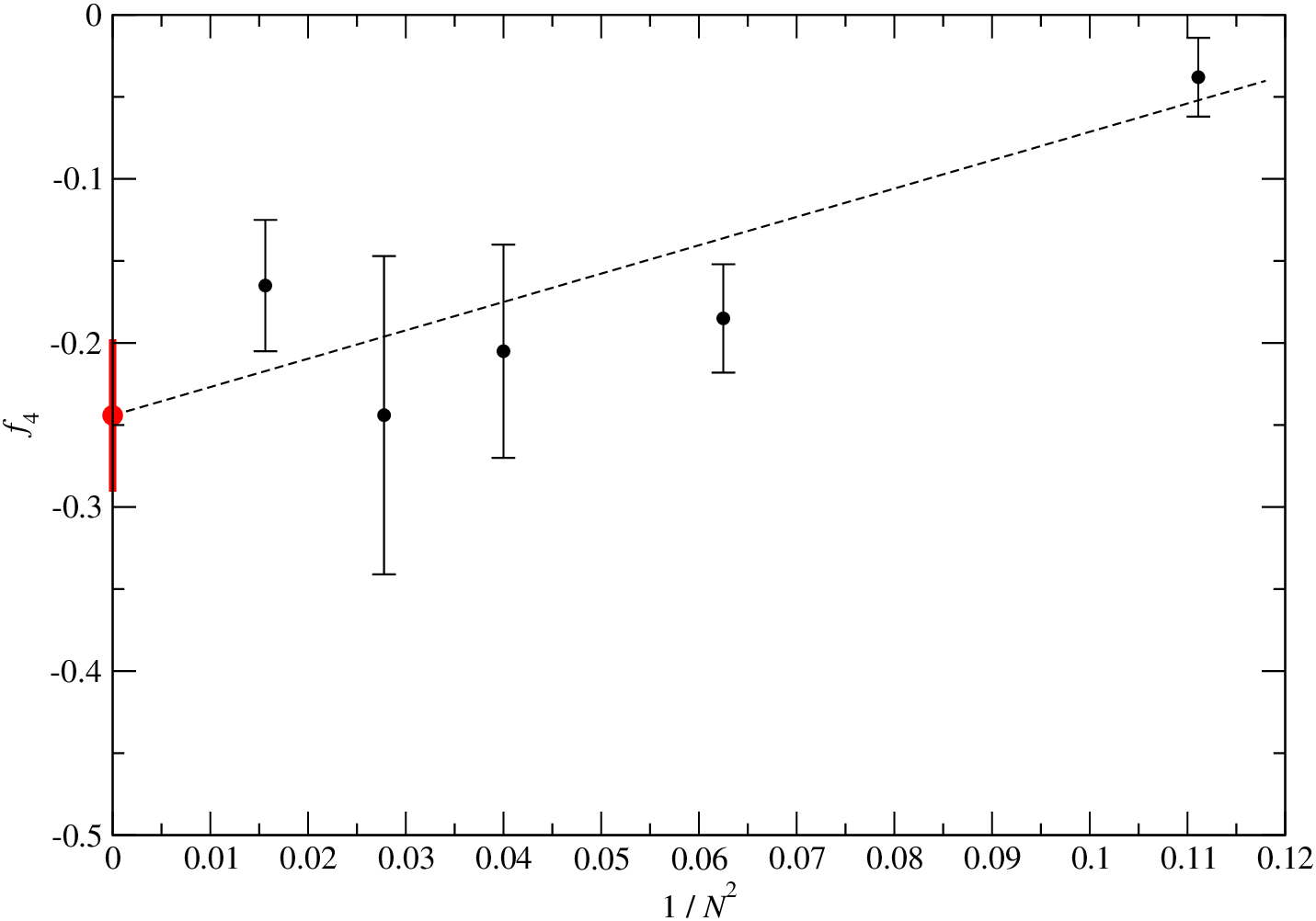}}
\caption{(Color online) Extrapolation of the parameters of eq.~(\protect\ref{interpolating_formula_for_Delta}), as obtained from fits to our data for the different $\SU(N)$ groups, to the large-$N$ limit. For each parameter, the result in the $N \to \infty$ limit is obtained through a first- (for $f_1$, $f_3$ and $f_4$) or second-order (for $f_2$) polynomial extrapolation in $N^{-2}$, according to eq.~(\protect\ref{large_N_parameter_extrapolation}).}
  \label{fig:Delta_parameters_large_N_extrapolation}
\end{figure}

The fit results are displayed in tab.~\ref{tab:Delta_parameters_large_N_extrapolation} and in fig.~\ref{fig:Delta_parameters_large_N_extrapolation}.

\begin{table}[h]
\centering
\begin{tabular}{|c||c|c|c||c|}  
\hline
parameter & $c_i$ & $l_i$ & $q_i$& $\chi^2/$d.o.f. \\
\hline \hline
 $f_1$ & $0.9918(20)$   & $0.361(21)$  & --               & $0.32$ \\
 $f_2$ & $0.00895(53)$ & $-0.217(16)$ & $3.30(11)$ & $0.01$ \\
 $f_3$ & $1.768(36)$     & $-0.66(41)$   & --               & $2.16$ \\
 $f_4$ & $-0.244(46)$    & $1.73(55)$    & --               & $1.56$ \\
\hline
\end{tabular}
\caption{Results of the fits for $\Delta$ according to eq.~(\protect\ref{interpolating_formula_for_Delta}); for $f_1$, $f_3$ and $f_4$, we perform a two-parameter fit, including the $c_i$ and $l_i$ terms in eq.~(\protect\ref{interpolating_formula_for_Delta}), setting $q_i=0$. However, this procedure would yield a non-physical, negative result for the $f_2$ extrapolation, therefore for this parameter we performed a three-parameter fit, fitting $c_i$, $l_i$, and also $q_i$. The errors shown do not include systematic uncertainties.}\label{tab:Delta_parameters_large_N_extrapolation}
\end{table}

\begin{figure}
\centerline{\includegraphics[width=.80\textwidth]{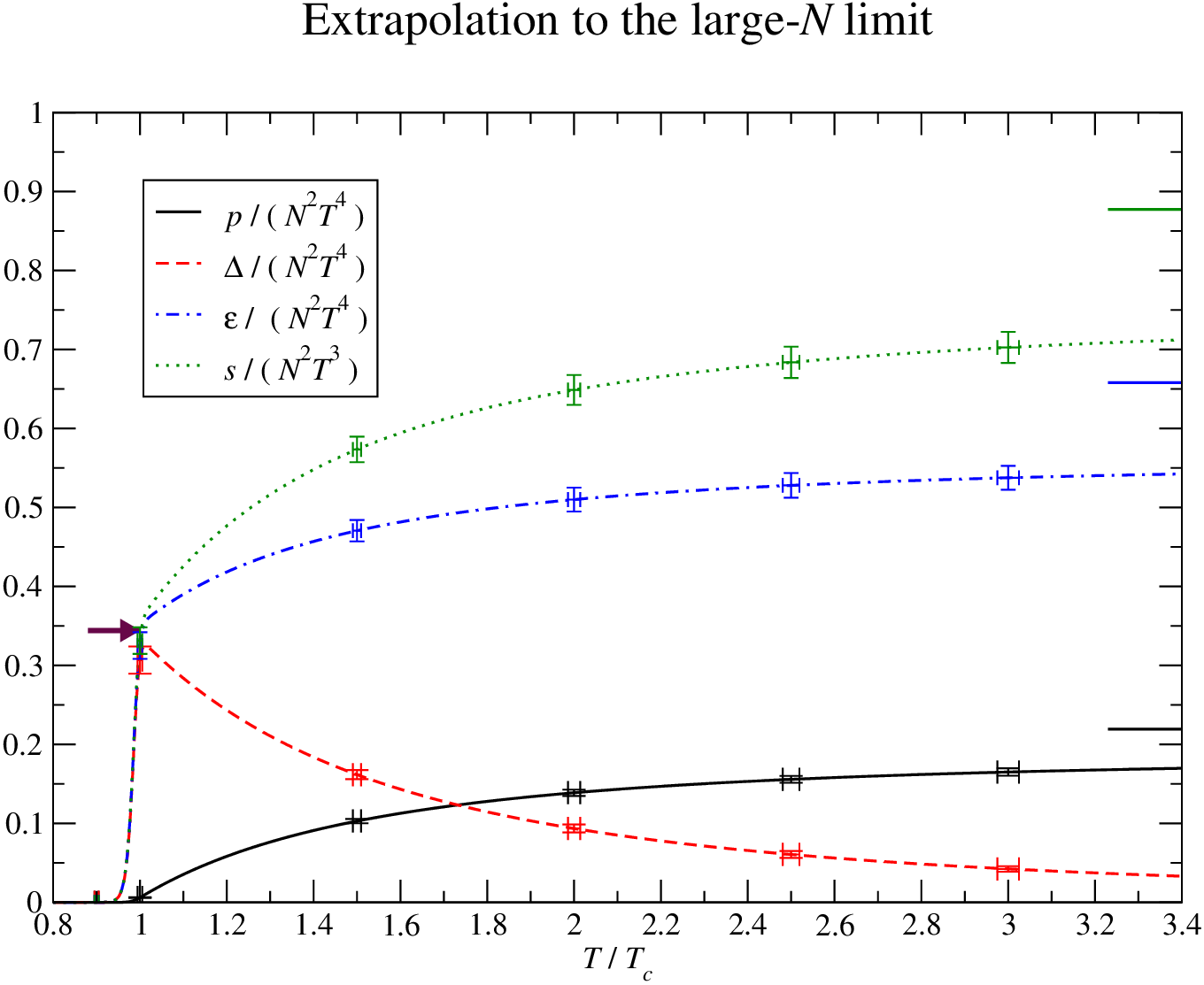}}
\caption{(Color online) Final results for the large-$N$ limits of $p/(N^2T^4)$ (black solid curve), $\Delta/(N^2T^4)$ (red dashed curve), $\epsilon/(N^2T^4)$ (blue dash-dotted curve) and $s/(N^2T^3)$ (green dotted curve); the errorbars (including statistical and systematic uncertainties) at six reference points corresponding to $T/T_c=0.9$, $1$, $1.5$, $2$, $2.5$ and $3$ are also shown. In the normalization used in this plot, the Stefan-Boltzmann limits for the pressure, energy density and entropy density are respectively denoted by the three horizontal bars displayed on the right-hand side of the plot, from bottom to top. The maroon arrow on the left-hand side of the figure denotes the large-$N$ limit of the latent heat $L_h$, as calculated in ref.~\protect\cite{Lucini:2005vg}, which is consistent with our result.}
  \label{fig:large_N_results}
\end{figure}

Finally, from the parameters thus extrapolated, we obtain the curves for the large-$N$ limit of $\Delta/(N^2 T^4)$, of $p/(N^2 T^4)$ (by numerical integration of the latter over $\log T$), and of $\epsilon/(N^2 T^4)$ and $s/(N^2 T^3)$ (through linear combinations of the former two quantities). The results are shown in fig.~\ref{fig:large_N_results} (note that the quantities in this plot are not normalized to their SB limits; the latter are denoted by the three horizontal bars on the right-hand side of the figure), where we also show the large-$N$ limit of the latent heat $L_h$  calculated in ref.~\cite{Lucini:2005vg}, namely: $L_h^{1/4}N^{-1/2}T_c^{-1}=0.766(40)$. Our estimate for the same quantity is fully consistent: $L_h^{1/4}N^{-1/2}T_c^{-1}=0.759(19)$. In this figure we also show our estimate of the errors at six reference temperatures $T/T_c=0.9$, $1$, $1.5$, $2$, $2.5$ and $3$.

\section{Conclusions and outlook}
\label{sec:conclusions}

In this article, we have presented the results of a lattice study of the main equilibrium thermodynamic observables in $\SU(N)$ gauge theories with $N=3$, $4$, $5$, $6$ and $8$ colors at finite temperature, and discussed their extrapolation to the large-$N$ limit. Our main motivation for this study was to investigate whether (and to what quantitative extent) the large-$N$ limit captures the non-trivial thermodynamics of the deconfined plasma in the theory with $N=3$ colors. This is crucially relevant for AdS/CFT models or holographic QCD models which are expected to provide an analytical theoretical description of the strongly-interacting QCD plasma produced in relativistic collisions of heavy ions. To address this issue, we have performed a high-precision numerical study, based on a fine temperature scan in the phenomenologically most interesting region between $0.8T_c$ and $3.4T_c$.

Our results, which agree with previous similar studies~\cite{Bringoltz:2005rr, Datta:2009tj}, show that the pressure, the trace of the stress-energy tensor, the energy density and the entropy density (per gluon) evaluated in all gauge theories that we considered are very close to each other and to those obtained in the $\SU(3)$ model. Essentially, the main difference between the results we obtained for $\SU(3)$ and for the other $\SU(N>3)$ groups is that, as expected on general grounds~\cite{Holland:2003kg, Holland:2003mc, Pepe:2004rc}, the latter exhibit a more strongly first-order deconfining phase transition.

This led us to compare our results with the improved holographic QCD model recently proposed in refs.~\cite{Gursoy:2007cb, Gursoy:2007er, Kiritsis:2009hu, Gursoy:2008bu}, which is an AdS/QCD model providing a quantitative description of the features of the strongly-coupled regime of the gauge theory in terms of a dual, five-dimensional gravity model with a dilaton potential depending on two free parameters. Our results turn out to be in very good agreement with the curves from the improved holographic QCD model published in ref.~\cite{Gursoy:2009jd}, as fig.~\ref{fig:pressure}, fig.~\ref{fig:rescaled_trace}, fig.~\ref{fig:energy} and fig.~\ref{fig:entropy} show. 

In the improved holographic QCD model (and in other similar models) the running of the coupling in the gauge theory is associated with a non-trivial dilaton in the dual gravity theory; in particular, the parameters of the dilaton potential can be chosen consistently with the quantitative results known for the gauge theory, both in the perturbative and in the non-perturbative regimes. This successfully reproduces the non-trivial features observed in the thermodynamic quantities considered in this work, including, in particular, their behavior in the temperature range above and close to the deconfinement temperature. 

The latter temperature range is indeed particularly interesting, not only from the  phenomenological and experimental point of view, but also from a theoretical perspective, since it is a regime where weak-coupling expansions of the pressure typically show large deviations from lattice results, and, in particular, fail to reproduce the large values of $\Delta$ observed in numerical simulations---see ref.~\cite{Hietanen:2008tv} and references therein. In particular, the observation of a possible $T^2$-dependence of $\Delta$ in the lattice results for $\SU(3)$ from  ref.~\cite{Boyd:1996bx} led to conjecture the existence of contributions which could not be captured by a perturbative approach~\cite{Pisarski:2006hz, Pisarski:2006yk}. Our results appear to confirm that the $T^2$-behavior $\Delta$ is common to all gauge groups that we studied, and thus is likely to also persists in the $N \to \infty$ limit (see fig.~\ref{fig:Pisarski}), but we cannot rule out the possibility that our data could also be fitted by some more complicated functional form.

Next, we discussed some issues related to the description of the QCD plasma in terms of AdS/CFT models, in particular the large-$N$ limit of the $\mathcal{N}=4$ supersymmetric Yang-Mills theory. While the gauge-gravity correspondence~\cite{Maldacena:1997re} provides one with formidable tools to study the strongly-coupled regime of this theory analytically, via calculations in the weakly-coupled regime of the dual supergravity theory, the supersymmetric theory is not a close approximation of real-world QCD; in particular, the former is conformally invariant, whereas in the latter the coupling runs with the momentum scale. However, it is possible that QCD may admit an effective AdS/CFT description, at least in a regime where it tends to be almost scale-invariant. In fig.~\ref{fig:pressure_versus_energy}, we have shown that, for temperatures around $3T_c$, the results for the pressure \emph{versus} the energy density for all gauge groups are already relatively close to the conformal limit $\epsilon=3p$, while the theory is still strongly interacting and far from the Stefan-Boltzmann limit. It is intriguing to observe that, if one identifies the physical 't~Hooft running coupling evaluated at such temperatures in the $\overline{\rm MS}$ scheme with the $\lambda$ parameter of the $\mathcal{N}=4$ model, then the supergravity prediction for the ratio of the entropy density to its limit in the non-interacting gas~\cite{Gubser:1998nz} is in very good agreement with the lattice results precisely in the expected temperature range, as fig.~\ref{fig:entropy_lambda} shows. Although, as we have discussed, such comparison between the Yang-Mills models and the supersymmetric theory may be questionable at a fundamental level, prone to scheme-dependence, and the observed agreement may be accidental (because the dependence of $s/s_0$ on $\lambda$ in the supergravity theory is only known at the first non-trivial order in an expansion around the strong-coupling limit), we also note that, if one just compared the lattice results with the $s/s_0$ value predicted by the supergravity model in the infinite-coupling limit, which is $3/4$~\cite{Gubser:1996de}, then in the temperature regime where the thermodynamic observables (in units of the appropriate powers of the temperature) tend to become almost $T$-independent the lattice results would overshoot the AdS/CFT prediction. This observation may be relevant to identify the parameters of AdS/CFT models appropriate to describe the QCD plasma~\cite{Noronha:2009vz}. 

Finally, we have discussed the systematic uncertainties affecting the calculations presented in this work (including, in particular, those related to the scale determination, to discretization effects, and to finite-volume effects), and performed an extrapolation to the large-$N$ limit of the various thermodynamic quantities; the results---for a parametrization of $\Delta$ according to eq.~(\ref{interpolating_formula_for_Delta})---are reported in the second column of  tab.~\ref{tab:Delta_parameters_large_N_extrapolation} and displayed in fig.~\ref{fig:large_N_results} (in which we also show the estimated errorbars at some reference temperatures).

To summarize, our results show that the non-trivial features of equilibrium thermodynamic observables in the strongly-interacting deconfined plasma are common to all gauge groups investigated in this work, and to the large-$N$ limit. Theoretical models based on the analytical simplifications and on the dualities that are conjectured to be realized in that limit are thus expected to be relevant for the thermodynamic properties of the plasma of $\SU(3)$ Yang-Mills theory (and, likely, of QCD) at finite temperature. This is particularly important to get analytical predictions about quantities (\emph{e.g.} transport coefficients, observables involving real-time dynamics) and/or regions of the QCD phase diagram (\emph{e.g.} at large densities) where lattice QCD calculations are hindered by technical difficulties.

As for possible future research lines related to this project, it would be interesting to  extend the present investigation of the thermodynamics of hot $\SU(N)$ gauge theories by looking at other observables, for instance screening masses, and comparing the results with predictions from the improved holographic QCD model~\cite{Gursoy:2007cb, Gursoy:2007er, Kiritsis:2009hu, Gursoy:2008bu} or from AdS/CFT~\cite{Bak:2007fk}. The investigation of observables related to non-equilibrium thermodynamics in the large-$N$ limit would also be interesting, but is expected to be much more demanding from the computational point of view. On the contrary, another interesting research direction would be to perform similar studies for models defined in lower dimensions~\cite{Bialas:2008rk}, and/or based on different gauge groups, not necessarily relevant for the large-$N$ limit, for which powerful numerical algorithms are available~\cite{Caselle:2002ah, Panero:2004zq, Gliozzi:2005ny, Panero:2005iu, Caselle:2006dv, Cherrington:2008ey, Cherrington:2008mf, Allais:2008bk}. This could lead to a better identification and understanding of the universal and of the model- or dimension-dependent features of strongly interacting systems at finite temperature, and possibly shed new light onto the fascinating mysteries of the QCD plasma.

\section*{Acknowledgements}

The author acknowledges support from the Alexander~von~Humboldt Foundation and from INFN, and thanks B.~Bringoltz, M.~Caselle, Ph.~de~Forcrand, J.~Erdmenger, M.~Fromm, F.~Gliozzi, U.~G\"ursoy, E.~Kiritsis, I.~Kirsch, A.~Kurkela, M.~Laine, B.~Lucini, F.~Nitti, G.~D.~Torrieri and P.~Weisz for correspondence, discussions and encouragement. The author also thanks the authors of ref.~\cite{Gursoy:2009jd} for providing the numerical values obtained in the improved holographic QCD model for various thermodynamic quantities, which are reproduced in figs.~\ref{fig:pressure}, \ref{fig:rescaled_trace}, \ref{fig:energy} and \ref{fig:entropy} of the present work. The University of Regensburg hosts the Collaborative Research Center SFB/TR 55 ``Hadron Physics from Lattice QCD''.

\end{document}